\documentclass{aa}
\usepackage{graphicx}
\usepackage{xcolor}
\usepackage{listings}
\usepackage{subfig}
\usepackage{threeparttable}

\usepackage[bookmarksnumbered=true]{hyperref}

\usepackage{scalerel}

\hypersetup{
     colorlinks = true,
     linkcolor = blue,
     anchorcolor = blue,
     citecolor = blue,
     filecolor = blue,
     urlcolor = blue
     }
\definecolor{dkgreen}{rgb}{0,0.6,0}
\definecolor{gray}{rgb}{0.5,0.5,0.5}
\definecolor{mauve}{rgb}{0.58,0,0.82}

\usepackage{txfonts}
\usepackage{cleveref}
\usepackage[LGRgreek]{mathastext}

\newcommand{\hdton}{HD~209458b\xspace}

\usepackage{txfonts}
\begin{document}

   \title{Retrieving the transmission spectrum of HD~209458b using CHOCOLATE: A new chromatic Doppler tomography technique\thanks{Based on Guaranteed Time Observations collected at the European Southern Observatory under ESO programme 1102.C-0744 by the ESPRESSO Consortium.}}

   \titlerunning{Retrieving transmission spectrum of HD~209458b using CHOCOLATE}
   
   \authorrunning{E.~Esparza-Borges, et al.}
   \author{E.~Esparza-Borges\inst{\ref{iiac},\ref{iull}},
          M.~Oshagh\inst{\ref{iiac},\ref{iull}},
          N.~Casasayas-Barris\inst{\ref{ileiden}},
          E.~Pall\'e\inst{\ref{iiac},\ref{iull}},
          G.~Chen\inst{\ref{ichina1},\ref{ichina2}},
          G.~Morello\inst{\ref{iiac},\ref{iull}},
          N.C.~Santos\inst{\ref{iporto1},\ref{iporto2}},
          J.V.~Seidel\inst{\ref{igeneve}},
          A.~Sozzetti\inst{\ref{iinaf_torino}},
          R.~Allart\inst{\ref{imontreal},\ref{igeneve}},
          P.~Figueira\inst{\ref{iesochile},\ref{iporto1}},
          V.~Bourrier\inst{\ref{igeneve}},
          J.~Lillo-Box\inst{\ref{icab}},
          F.~Borsa\inst{\ref{iinaf_brera}},
          M.R.~Zapatero Osorio\inst{\ref{icab2}},
          H.~Tabernero\inst{\ref{iporto1},\ref{iporto2},\ref{icab2}},
          O.D.S.~Demangeon\inst{\ref{iporto1},\ref{iporto2}},
          V.~Adibekyan\inst{\ref{iporto1},\ref{iporto2}},
          J.I.~González Hernández\inst{\ref{iiac},\ref{iull}},
          A.~Mehner\inst{\ref{iesochile}},
          C.~Allende Prieto\inst{\ref{iiac},\ref{iull}},
          P.~Di Marcantonio\inst{\ref{iinaf_trieste}},
          Y.~Alibert\inst{\ref{ibern}},
          S.~Cristiani\inst{\ref{iinaf_trieste}},
          G.~Lo Curto\inst{\ref{iesochile}},
          C.J.A.P.~Martins\inst{\ref{iporto1},\ref{iporto3}},
          G.~Micela\inst{\ref{iinaf_palermo}},
          F.~Pepe\inst{\ref{igeneve2}},
          R.~Rebolo\inst{\ref{iiac},\ref{iull},\ref{icsic}},
          S.G.~Sousa\inst{\ref{iporto1},\ref{iporto2}},
          A.~Suárez Mascareño\inst{\ref{iiac},\ref{iull}}
          and
          S.~Udry\inst{\ref{igeneve2}}}

\institute{
\label{iiac} Instituto de Astrof\'isica de Canarias, E-38200 La Laguna, Tenerife, Spain
\and
\label{iull} Departamento de Astrof\'isica, Universidad de La Laguna, E-38206 La Laguna, Tenerife, Spain
\and
\label{ileiden} Leiden Observatory, Leiden University, Postbus 9513, 2300 RA Leiden, The Netherlands
\and
\label{ichina1} CAS Key Laboratory of Planetary Sciences, Purple Mountain Observatory, Chinese Academy of Sciences, Nanjing 210023, People's Republic of China
\and
\label{ichina2} CAS Center for Excellence in Comparative Planetology, Hefei 230026, People's Republic of China
\and
\label{iporto1} Instituto de Astrofísica e Ciências do Espaço, Universidade do Porto CAUP, Rua das Estrelas 4150-762 Porto, Portugal
\and
\label{iporto2} Departamento de Física e Astronomia Faculdade de Ciências, Universidade do Porto, Rua do Campo Alegre 687, PT4169-007 Porto, Portugal
\and
\label{iporto3} Centro de Astrof\'{\i}sica da Universidade do Porto, Rua das Estrelas, 4150-762 Porto, Portugal
\and
\label{igeneve} Observatoire astronomique de l’Universit\'e de Gen\`eve, Chemin Pegasi 51, 1290 Versoix, Switzerland
\and
\label{igeneve2} Université de Gen\`eve, Observatoire Astronomique, Chemin Pegasi 51, 1290 Versoix, Switzerland
\and
\label{iinaf_torino} INAF – Osservatorio Astrofisico di Torino, Via Osservatorio 20, I- 10025 Pino Torinese, Italy
\and
\label{iinaf_trieste} INAF – Osservatorio Astronomico di Trieste, via G. B. Tiepolo 11, I-34143, Trieste, Italy
\and
\label{iinaf_brera} INAF – Osservatorio Astronomico di Brera, Via E. Bianchi 46, 23807 Merate (LC), Italy
\and
\label{iinaf_palermo} INAF – Osservatorio Astronomico di Palermo, Piazza del Parlamento 1, 90134 Palermo, Italy
\and
\label{imontreal} Department of Physics, and Institute for Research on Exoplanets, Universit\'e de Montr\'eal, Montr\'eal, H3T 1J4, Canada
\and
\label{iesochile} European Southern Observatory, Alonso de C\'ordova 3107, Vitacura, Región Metropolitana, Chile
\and
\label{icab} Centro de Astrobiolog\'ia (CSIC-INTA), Depto. de Astrof\'isica, ESAC campus 28692 Villanueva de la Cañada (Madrid), Spain
\and
\label{icab2} Centro de Astrobiolog\'ia (CSIC-INTA), Carretera de Ajalvir km 4, E-28850 Torrej\'on de Ardoz, Madrid, Spain
\and
\label{ibern} Physics Institute, University of Bern, Sidlerstrasse 5, 3012 Bern, Switzerland
\and
\label{iifundtrieste} Institute for Fundamental Physics of the Universe, Via Beirut 2, I-34151 Grignano, Trieste, Italy
\and
\label{ilisboa1} Instituto de Astrofísica e Ciências do Espaço, Faculdade de Ciências da Universidade de Lisboa, Campo Grande, PT1749-016
Lisboa, Portugal
\and
\label{ilisboa2} Faculdade de Ciências da Universidade de Lisboa (Departamento de F\'isica), Edif\'icio C8, 1749-016 Lisboa, Portugal
\and
\label{icsic} Consejo Superior de Investigaciones Cientif\'icas, Spain
}

   \date{Received September 15, 1996; accepted March 16, 1997}

 
  \abstract{Multiband photometric transit observations or low-resolution spectroscopy (spectro-photometry) are normally used to retrieve the broadband transmission spectra of transiting exoplanets in order to assess the chemical composition of their atmospheres. In this paper, we present an alternative approach for recovering the broadband transmission spectra using chromatic Doppler tomography based on physical modeling through the \texttt{SOAP} tool: CHOCOLATE (CHrOmatiC  line prOfiLe  tomogrAphy TEchnique). To validate the method and examine its performance, we use observational data recently obtained with the ESPRESSO instrument to retrieve the transmission spectra of the archetypal hot Jupiter HD~209458b. Our findings indicate that the recovered transmission spectrum is in good agreement with the results presented in previous studies, which used different methodologies to extract the spectrum, achieving similar precision. We explored several atmospheric models and inferred from spectral retrieval that a model containing H$_2$O and NH$_3$ is the preferred scenario. The CHOCOLATE methodology is particularly interesting for future studies of exoplanets around young and active stars or moderate to fast rotating stars, considering \texttt{SOAP}'s ability to model stellar active regions and the fact that the rotational broadening of spectral lines favors its application. Furthermore, CHOCOLATE will allow the broad transmission spectrum of a planet to be retrieved using high S/N, high-resolution spectroscopy with the next generation of Extremely Large Telescopes (ELTs), where low-resolution spectroscopy will not always be accessible.}


   \maketitle
%

\section{Introduction}

During its passage in front of its rotating host star, a transiting exoplanet creates a radial velocity (RV) signal, which is generated by obstructing the Doppler-shifted components of the portion of the stellar disk that is blocked by the planet. This is known as the Rossiter–McLaughlin (RM) effect \citep{Holt-1893, Rossiter-24, McLaughlin-24}. The RM signal contains several important pieces of information, including the sky-projected planetary spin-orbit angle \cite[and references therein]{Triaud2018_HbOE}. Similar to the depth of the photometric transit light curve, the RM semi-amplitude also scales with the square of the planet-star radius ratio \citep{Snellen-04}. Recently, several studies employed the chromatic RM technique using HARPS, CARMENES and ESPRESSO spectrograph data on HD~189733 and HD~209458  \citep{DiGloria-15, Oshagh-20, Santos-20}, and they showed that it can be a powerful method for probing wide broadband features, which are challenging to probe from the ground. Moreover, \citet{Oshagh-20} demonstrated the advantage of chromatic RM analysis by combining HARPS and CARMENES observations, which covered a wide wavelength range and allowed precise estimations of the stellar active regions' properties and the mitigation of their impact on the retrieved transmission spectra.

During a transit, the planet occults different patches of the stellar surface along its trajectory, effectively removing the corresponding contribution to the stellar line profile at that local stellar velocity. Thus, the traveling Doppler shadow cast by the planet creates an identifiable distortion in the line profiles. Assessing these deformations is called line-profile tomography. This method has been very useful for successfully estimating the sky-projected planetary spin-orbit angle for planets around fast rotating host stars \citep[e.g.,][]{CollierCameron-10, Hartman-15, Zhou-17, Johnson-15}. As for the transit's depth in photometric transit light curves, the line-profile distortions also scale with the planet-star radius ratio.
Thus, measuring line-profile distortions over several wavelength bins yields a low-resolution broadband transmission
spectrum of the planet atmosphere.

The chromatic RM is based on the strong dependence of the RM signal's amplitude on the planet-star radius ratio. Thus, the study of this dependence at different wavelength ranges leads to the determination of the planetary radius as a function of wavelength. Currently, \cite{DiGloria-15}, \cite{Oshagh-20} and \cite{Santos-20} have been the only studies to make use of the chromatic RM technique to retrieve transmission spectra of HD~189733b and HD~209458b.

\citet{Borsa_2016} attempted, for the first time, the chromatic line-profile tomography technique on the well-known exoplanet HD~189733b using HARPS observations, obtaining results broadly consistent with other ground- and space-based observations. However, their preprocessing steps and their modeling approach were criticized by \citet{Cegla-17}, who showed that it is not possible to obtain exoplanet radius measurements with ground-based observations using the methods applied in \cite{Borsa_2016} due to the transit light curve normalization necessary to remove the effects of the Earth’s atmosphere. The approximate model used to calculate the planet radius from the deformations in the chromatic line profiles was also called into question.

Here we present an alternative method for retrieving the transmission spectrum from chromatic line-profile tomography based on physical modeling, and we validate its performance using ESPRESSO spectra of HD~209458b. \hdton is a hot Jupiter orbiting around a solar-type star. It was the first exoplanet detected by the transit technique \citep{Charbonneau2000,Henry2000} and the first exoplanet whose atmosphere was detected \citep{Charbonneau2002}. Since then, several spectral features (e.g., Na, $H_{2}O$, CO, He, TiO, $CH_{4}$, HCN, NH3 and Fe+) have been detected in the atmosphere of \hdton \citep{Snellen2010,Sing-16,Tsiaras2016,Tsiaras2018,AlonsoFloriano2019,Brogi2019,SanchezLopez2019,Cubillos2020,Santos-20,Giacobbe2021}. In Sect.~\ref{sec:Observation}  we present our observations and data reduction process. In Sects.~\ref{sec:Model} and \ref{sec:Fitting} we present the details of our chromatic Doppler tomography methodology and its implementation, describing the modeling and fitting procedures, respectively. In Sect.~\ref{sec:Results} the results of our analysis and a spectral retrieval on them are presented, along with the validation and examination of whether the result obtained by \citet{Santos-20} can be reproduced using the same ESPRESSO observations.We discuss and conclude our study in Sect.~\ref{sec:Conclusion}.

\section{Observations and methodology}\label{sec:Observation}
\subsection{ESPRESSO observations}
Two transits of HD~209458b were observed with ESPRESSO \citep{Pepe-12, Pepe-21}, a fiber-fed spectrograph located at the Very Large Telescope (VLT) that covers a wavelength range from 380~nm to 788~nm at a resolution of R $\sim$ 140000, on the nights of July 19, 2019 and September 10, 2019. The observations were acquired through the ESPRESSO consortium's guaranteed time observations (GTO) under program 1102.C-0744 and had already been analyzed using different methodologies by \cite{Santos-20} and \cite{CasasayasBarris_2021}. Both transits were observed using the UT3 telescope in the HR21 mode with the same exposure time of 175 seconds, which yielded a total of 89 and 85 exposures with an averaged signal-to-noise ratio (S/N) of 234 and 193 at 588~nm (physical order 104) for the two nights, respectively. It should be noted that the HR21 mode employs a slower readout that introduces a lower noise value per readout, and reads two pixels in the spatial direction at once to reduce the number of readouts. The resulting spectra were reduced using the latest Data Reduction Software (DRS) pipeline version 2.2.8. Each observed spectrum was divided into 85 echelle interference “orders”. The DRS calculates and delivers the cross-correlation function (CCF) for each spectral order, as well as for the whole wavelength range. Based on the spectral type of the host star, HD~209458, the CCFs were generated using an F9V mask. We note that some orders are heavily contaminated by tellurics, and thus they are masked by the DRS and have no CCFs available. In this work we present the results obtained using the sky-subtracted (SKYSUB) data set. Nevertheless, we see no significant impact on the final results when using the SKYSUB or the non-SKYSUB products extracted by the DRS (see Appendix~\ref{App:SKYSUB_comparison}). Additionally, due to the optical design of ESPRESSO and its image slicer, each order was repeated twice, and thus in practice we see a total of 170 “slices” \citep{Pepe-21, Santos-20}.

\subsection{Data analysis}\label{sec:data_reduction}
We processed our observations by applying the following procedure:

\begin{enumerate}

\item Normalization of CCFs: The continuum levels of the CCFs are arbitrary because ESPRESSO observations are not flux-calibrated. For this reason, we needed to normalize each CCF by its continuum level. The normalization was done by fitting a second-order polynomial\footnote{We chose the second-order polynomial for CCF normalization after testing both first- and second-order choices and concluding that the second-order polynomial allowed for a more robust fitting to the continuum, allowing for better CCF normalization with less residuals in the continuum.} to the continuum of each CCF. Specifically, we defined the continuum as regions with -20 < v < -10 km\,s$^{-1}$ and 11 < v < 14 km\,s$^{-1}$ of each individual CCF in each slice (i.e. the continuum regions marked by the solid gray lines in Fig.~\ref{masterccfs}). The continuum regions are defined asymmetrically to avoid some anomalies observed in the right side of the CCF continuum for certain slices\footnote{These anomalies may be related to the humps caused by blended lines described in \cite{lafarga2020carmenesccf}.}. Subsequently, we normalized each CCF by dividing them by the fitted polynomial. 

\item Subtraction of the Keplerian motion of the star: The objective of this step was to remove the stellar RV that is induced by the planet while leaving the RM signal unaffected. Each CCF was shifted to the stellar rest frame via the introduction of an offset in its RV. This allowed the Keplerian motion of the star induced by its planet to be subtracted, leaving only the imprint in RV due to the rotation of the star (in addition to the instrumental broadening). The systemic velocity was removed as well. To perform this step, we used \textit{SinRadVel} function from the \emph{PyAstronomy} python module \citep{PyAstronomy2019}, which calculates and returns the RV shift according to the defined planetary and stellar parameters: orbital period, mid-transit time, stellar RV semi-amplitude, and the host star's systemic RV (see Table~\ref{tab:systemparameters}).
Finally, all CCFs were linearly interpolated to the same RV grid (defined in the $-20 < v < 20 km\,s^{-1}$ range with a step of $0.5~km,s^{-1}$) so that every CCF is evaluated at exactly the same RV values. The interpolation was required in order to subsequently perform mathematical operations between CCFs.

\item Definition of the wavelength bins: The CCFs were reassembled in eight-slice bins using the mean of the slices.  
The binning definition applied here is analogous to the one presented in \cite{Santos-20}. The exact correspondence between the bins and the range of wavelengths covered by each bin is given in Table~\ref{tab:wavelength_correspondance_and_results}. We note that there is a slight overlap in wavelength between adjacent bins as a consequence of the way orders are created by dispersion plus cross-dispersion, as seen on ESPRESSO and similar echelle spectrographs.

\item Generation of the master CCF: We calculated the average of all the out-of-transit CCFs (master CCFs hereafter) for each wavelength
bin. The master CCF of each bin for Night 1 is presented in Fig.~\ref{masterccfs}.

\begin{figure}[h]
    \centering
    \includegraphics[width=1.\linewidth]{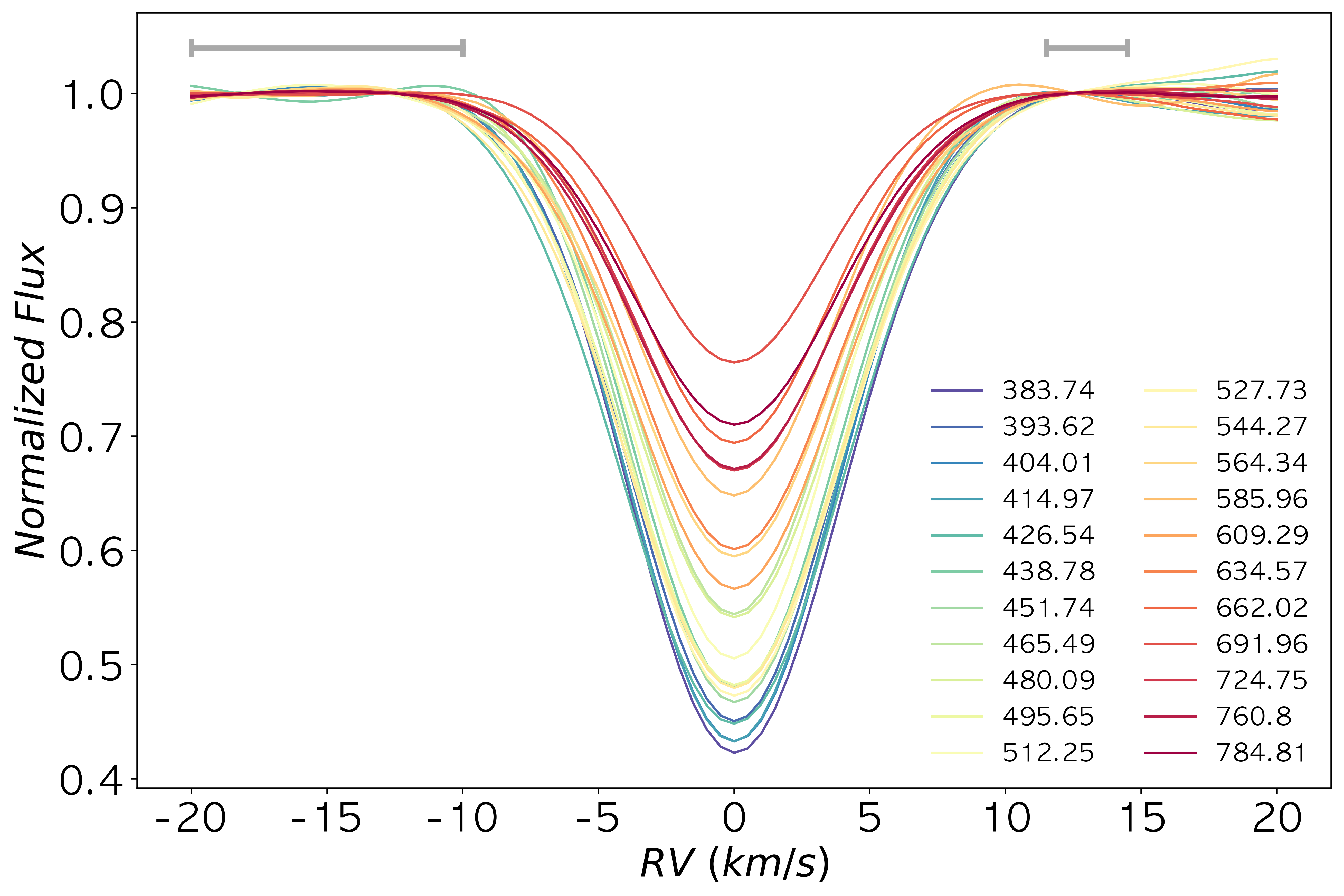}
    \caption{Master CCF of each bin (central wavelength of each bin given in nm) from the Night 1 eight-slice-bin data sets. The solid gray lines mark the continuum regions of the CCFs that were used in the normalization.}
    \label{masterccfs}
\end{figure}

\item Subtraction of the master CCF: For each spectral bin, we subtracted its master CCF from the whole CCF times series. This subtraction yielded a chronological map of residual CCFs, showing the Doppler tomography anomalies expected during transit (see Fig.~\ref{processed_DT_Map}).

\begin{figure}[hbt]
    \centering
    \includegraphics[width=1.\linewidth]{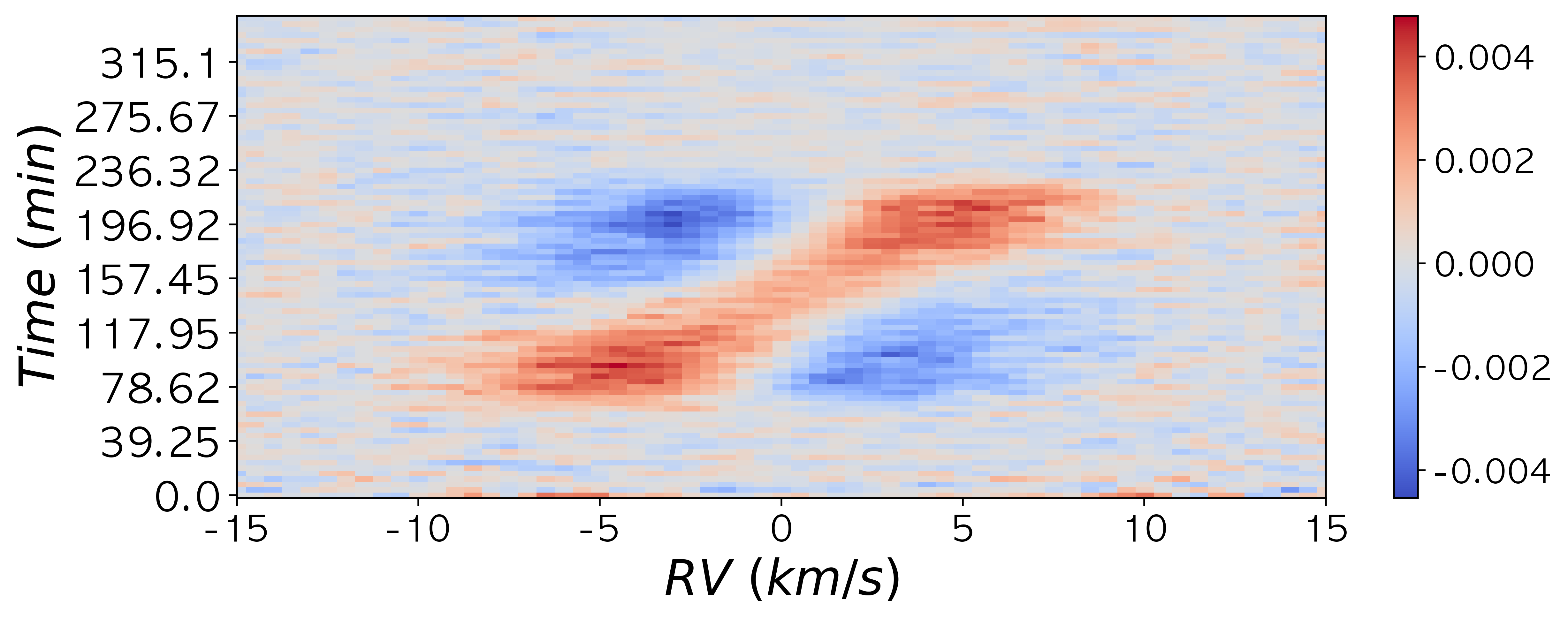}
    \includegraphics[width=1.\linewidth]{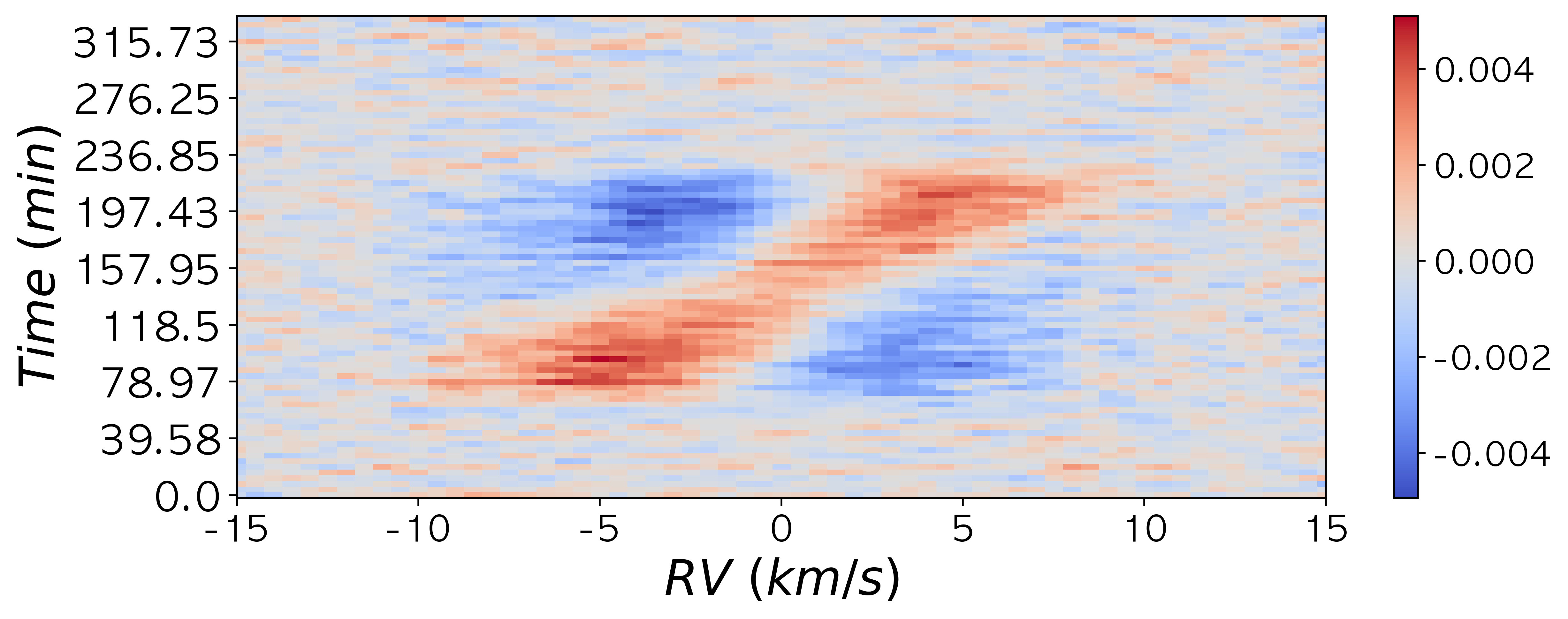}
    \caption{CCF residual matrix using the 439-nm-centered bin from Night 1 (top) and Night 2 (bottom) after applying the steps from Sect.~\ref{sec:data_reduction}.}
    \label{processed_DT_Map}
\end{figure}

Following the application of the above steps to each data set, a chronological sequence of residual CCFs for each bin was obtained, what we refer to as "CCF residual matrix". Figure~\ref{ResidualCCFs_examples} shows five residual CCFs (out-of-transit, in-transit, and mid-transit) for the 439-nm-centered bin of the ESPRESSO Night~1 data set to demonstrate the data reduction procedure. It is worth noting that each residual CCF in Fig.~\ref{ResidualCCFs_examples} corresponds to one row in the maps shown in Fig.~\ref{processed_DT_Map}.  Finally, a Python function was developed to aggregate the data collected over different nights and compute the average of the residual CCF maps. We note that the observations were acquired using the same exposure time for both nights (175~seconds), so the number of samples in-transit are exactly the same for both nights. Thus, in practice we combined data points that have very similar in-transit phases.

\begin{figure}[hbt]
    \centering
    \includegraphics[width=1.\linewidth]{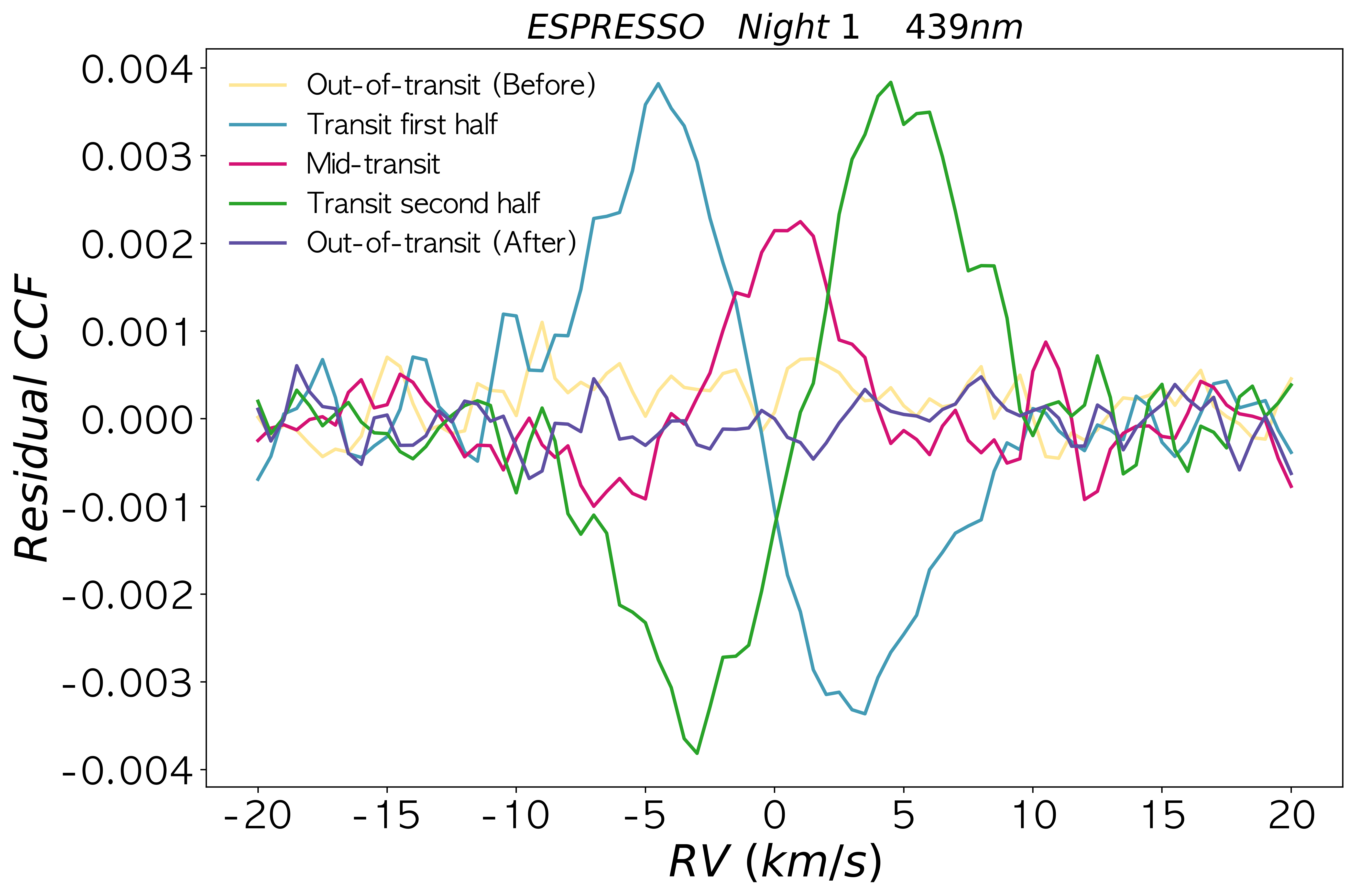}
    \caption{Residual CCFs from the ESPRESSO Night 1 data set at the 439-nm bin. Two out-of-transit CCFs (before and after transit) and three in-transit CCFs (mid-transit and before and after mid-transit) are shown.}
    \label{ResidualCCFs_examples}
\end{figure}

\item Error estimation: The noise of each data set in each wavelength bin was estimated with the standard deviation calculated at the RV regions in the CCF residual matrix that were far from the anomalies observed during the transit (regions marked as line-filled rectangles in Fig.~\ref{noise_example}).

\end{enumerate}

\section{Model}\label{sec:Model}

To model our observed CCF residual matrix we used the publicly available tool \texttt{SOAP3.0} (hereafter \texttt{SOAP}). This tool uses a pixelation approach to simulate a transiting
planet in front of a rotating host star and delivers the overall photometric and RV measurements of the system \citep{Boisse-12, Oshagh-13a, Dumusque-14, Akinsanmi-18}. Specifically, \texttt{SOAP} divides the star into a grid of $N \times N$ cells, each of which has its own Gaussian, that is defined by full width at half maximum (FWHM) $\sigma_{0}$ and depth parameters $prof_{0}$, and corresponds to the non-rotating star’s CCF. Subsequently, each cell's Gaussian is Doppler-shifted to account for stellar rotation, and its intensity is weighted using a quadratic limb-darkening law. Finally, the intensity in each cell is reduced to zero if the cell is blocked by the transiting planet.
All of the cells are eventually summed up to obtain a single integrated CCF over the entire stellar disk. In order to have a model compatible with our observed CCF residual matrix, steps similar to fourth and fifth steps in the data reduction procedure were carried out on the \texttt{SOAP} output CCFs. This led to a \texttt{SOAP} CCF residual matrix which can be used to model the observations.

The coefficients of the quadratic stellar limb-darkening law ($u_{1}$ and $u_{2}$), which are the only wavelength-dependent \texttt{SOAP} parameters, are taken from the Limb Darkening Toolkit (\texttt{LDTk}) estimations \citep{Parviainen-15}. \texttt{LDTk} provides an estimation of the quadratic limb-darkening coefficients for each wavelength bin based on the effective temperature (\textit{$T_{eff}=6118 \pm 25~K$}), the surface gravity (\textit{$log~g=4.36 \pm 0.04$}), and the metallicity (\textit{$[Fe/H]=0.02 \pm 0.05~dex$}) of the star \citep{Sousa_2008}.

\section{Fitting procedure}\label{sec:Fitting}

As previously mentioned, among all the fixed parameters in the \texttt{SOAP} models, the FWHM and depth values of the Gaussian-modeled CCF are especially relevant to performing a chromatic fitting to the data using \texttt{SOAP} models. For this reason, prior to the radius and limb-darkening coefficients fitting process (explained in Sect.~\ref{sec:rad_limb_fitting}), we needed to fit the \texttt{SOAP} CCF to the master CCF in each wavelength bin through a minimum $\chi^{2}$ method (Fig.~\ref{fig: sigmaprof_fit}) and obtain the FWHM and depth values of the Gaussian from the best fit. The resulting values for the FWHM and the depth of the Gaussian were fixed in \texttt{SOAP} configuration for each wavelength bin\footnote{These two parameters could be fit as extra free parameters through the \texttt{emcee} fitting procedure explained in Sect.~\ref{sec:rad_limb_fitting}. However, after testing this in several wavelength bins, we discarded the five-free-parameter fitting option because it implied a high computational cost without a significant contribution to the final results.} before the \texttt{emcee}-based fitting procedure was executed to determine the $R_{P}/R_{*}$, $u_{1}$, and $u_{2}$ posterior distributions. We are aware that the estimated values of $R_{P}/R_{*}$ might be affected by the choice of $P_{rot}$. However, since we are only interested in the relative changes in $R_{P}/R_{*}$ as a function of wavelength, this arbitrary offset will not lead to any misinterpretation of the retrieved transmission spectra.

\subsection{Radius and limb-darkening coefficients fitting}\label{sec:rad_limb_fitting}
Our fitting procedure is based on fitting the observed chromatic CCF residual matrix with a synthetic CCF residual matrix generated with \texttt{SOAP} \footnote{We only considered in-transit regions of CCF residual matrix in our fitting procedure.}. We used the planet radius and limb-darkening coefficients as our free parameters during the fitting process since they are the only wavelength-dependent parameters. The remaining required parameters in \texttt{SOAP} were adjusted to their values for HD~209458 and HD~209458b, which were taken from \cite{Mazeh2000}, \cite{Winn_2005_ip}, \cite{Southworth10}, \cite{Wang_Ford_2011}, \cite{Boyajian15}, \cite{Bonomo17} and \cite{Santos-20}, and are listed in Table~\ref{tab:systemparameters}. Since we are only interested in the relative changes in planet radius as a function of wavelength, fitting these fixed \texttt{SOAP} parameters is not necessary. Thus, we only fit $R_{P}/R_{*}$, $u_{1}$ and $u_{2}$ through this procedure. This is a common approach for retrieving transmission spectra frequently used in the chromatic RM technique \citep{DiGloria-15, Oshagh-20, Santos-20}.

\begin{figure*}[htb]
    \centering
    \includegraphics[width=1.\linewidth]{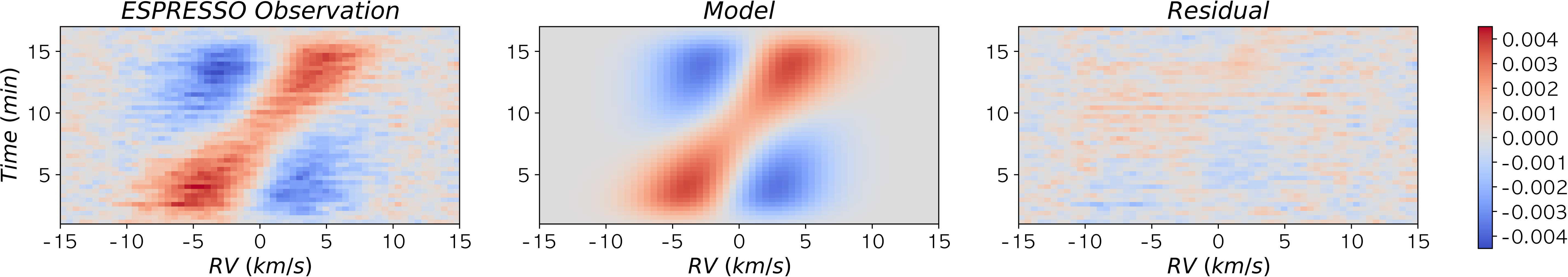}
    \caption{Comparison between the observed CCF residual matrix for the 439-nm-centered bin (left), the best-fit \texttt{SOAP}-modeled CCF residual matrix (center) and the residuals of the fitting (right).}
    \label{fig: model_obs_res}
\end{figure*}

\begin{table}[htb]
    \centering
    \caption{Stellar, planetary, and simulation parameters fixed in \texttt{SOAP} configuration to produce \texttt{SOAP} models.}
    \begin{tabular}{l r r}
        \hline \hline 
        \noalign{\smallskip}
        Parameter & Unit & Value \\
        \noalign{\smallskip}
              \hline 
       \noalign{\smallskip}
        \multicolumn{3}{c}{\dotfill Stellar parameters\dotfill}\\
        \noalign{\smallskip}
        Radius (\textit{$R_{*}$}) $^{(a)}$ & [\textit{$R_{\odot}$}] & 1.203\\
        Rotation Period (\textit{$P_{rot}$}) $^{(b)}$ & [\textit{days}] & 15.7\\
        Inclination Angle (\textit{i}) $^{(c)}$ & [\textit{º}] & 90\\
        Effective Temperature (\textit{$T_{eff}$}) $^{(d)}$ & [\textit{K}] & 6118\\
    \noalign{\smallskip}
                \multicolumn{3}{c}{\dotfill Planet parameters\dotfill}\\
                \noalign{\smallskip}
        Orbital Period (\textit{$P_{orb}$}) $^{(e)}$ & [\textit{days}] & 3.52472\\
        Eccentricity (\textit{e}) $^{(f)}$ &  & 0.0082\\
        Argument of Periastron (\textit{$\omega$}) & [\textit{º}] & 90\\
        Inclination of the Planetary Orbit (\textit{$i_{p}$}) $^{(g)}$ & [\textit{º}] & 86.59\\
        Projected Spin-orbit Angle ($\lambda$) $^{(h)}$ & [\textit{º}] & 0.6\\
        Semi-major axis (\textit{a}) $^{(c)}$ & [\textit{$R_{*}$}] & 8.48509\\
        \noalign{\smallskip}
        \multicolumn{3}{c}{\dotfill \texttt{SOAP} simulation parameters\dotfill}\\
        \noalign{\smallskip}
        Grid & & 300\\
        Instrument Resolution & & 100000\\
        RV step of CCF & [\textit{km\,s$^{-1}$}] & 0.5\\
        Width of CCF window & [\textit{km\,s$^{-1}$}] & 25\\
        \hline
    \end{tabular}
    \begin{tablenotes}
        \small
        \item $^{(a)}$ \cite{Boyajian15}  $^{(b)}$ \cite{Mazeh2000}  $^{(c)}$ \cite{Southworth10}  $^{(d)}$ \cite{Sousa_2008}  $^{(e)}$ \cite{Bonomo17}  $^{(f)}$ \cite{Wang_Ford_2011}  $^{(g)}$ \cite{Winn_2005_ip}  $^{(h)}$ \cite{Santos-20}
    \end{tablenotes}
    \label{tab:systemparameters}
\end{table}

Applying a Markov chain Monte Carlo (MCMC) approach, using the affine invariant ensemble sampler \texttt{emcee} package \citep{Foreman-Mackey-13}, the best-fit parameters and their associated uncertainties were calculated. The initial values for our free parameters were randomly initiated for ten MCMC chains inside the prior distributions. For each chain, we used a burn-in phase of 300 steps, and then again sampled
the chains for 1000 steps. Thus, the results concatenated to produce 10000 steps. The best-fit values were estimated by calculating the median values of the posterior distributions for each parameter. 

Because the estimates of the limb-darkening coefficients were calculated using a stellar model, they may be inaccurate. It has been shown that incorrect values can induce biases in the retrieved transmission spectra in multiband photometry \citep[e.g.][]{Csizmadia-13,Espinoza_2015,Espinoza_2016,Morello_2017,Morello_2018}. Thus, we left the $u_{1}$ and $u_{2}$ coefficients free during the fitting procedure\footnote{Our estimated limb-darkening coefficients values are close to the LDTk values; therefore, fixing the limb-darkening coefficients would not influence the retrieved transmission spectrum.}.

We imposed a very wide and uninformative uniform prior on the planet radius, $0.1<R_{P}/R_{*}<0.2$, which is 100 times wider than the prior imposed by \cite{Santos-20}. The priors on the limb-darkening coefficients were constrained by Gaussian priors created using \texttt{LDTk} \citep{Parviainen-15} for all the wavelength bins (i.e., $\mathcal{N}(\mu_{\texttt{LDTk}};0.05)$\footnote{$\mathcal{U}(a;b)$ is a uniform prior with lower and upper limits of $a$ and $b$. $\mathcal{N}(\mu; \sigma)$ is a normal distribution with mean $\mu$ and width $\sigma$.}). We note that 0.05 corresponds to three times the limb-darkening coefficients errors estimated by \texttt{LDTk}.

To illustrate the performance of our fitting procedure, we show in Fig.~\ref{fig: model_obs_res} a comparison between an observed CCF residual matrix, its best-fit \texttt{SOAP}-modeled CCF residual matrix, and the residuals between the best-fit model and observations. This \texttt{emcee}-based fitting approach allows the \emph{posterior} distributions of the planet-to-star radius ratio and the limb-darkening coefficients to be obtained (Fig.~\ref{CornerPlotExample}) at every defined wavelength bin. In this way, we determined the values of these parameters and their uncertainties as the most probable value and the $1\sigma$ error in each posterior distribution.

\section{Results}\label{sec:Results}

\begin{figure}[h]
    \centering
    \includegraphics[width=1.\linewidth]{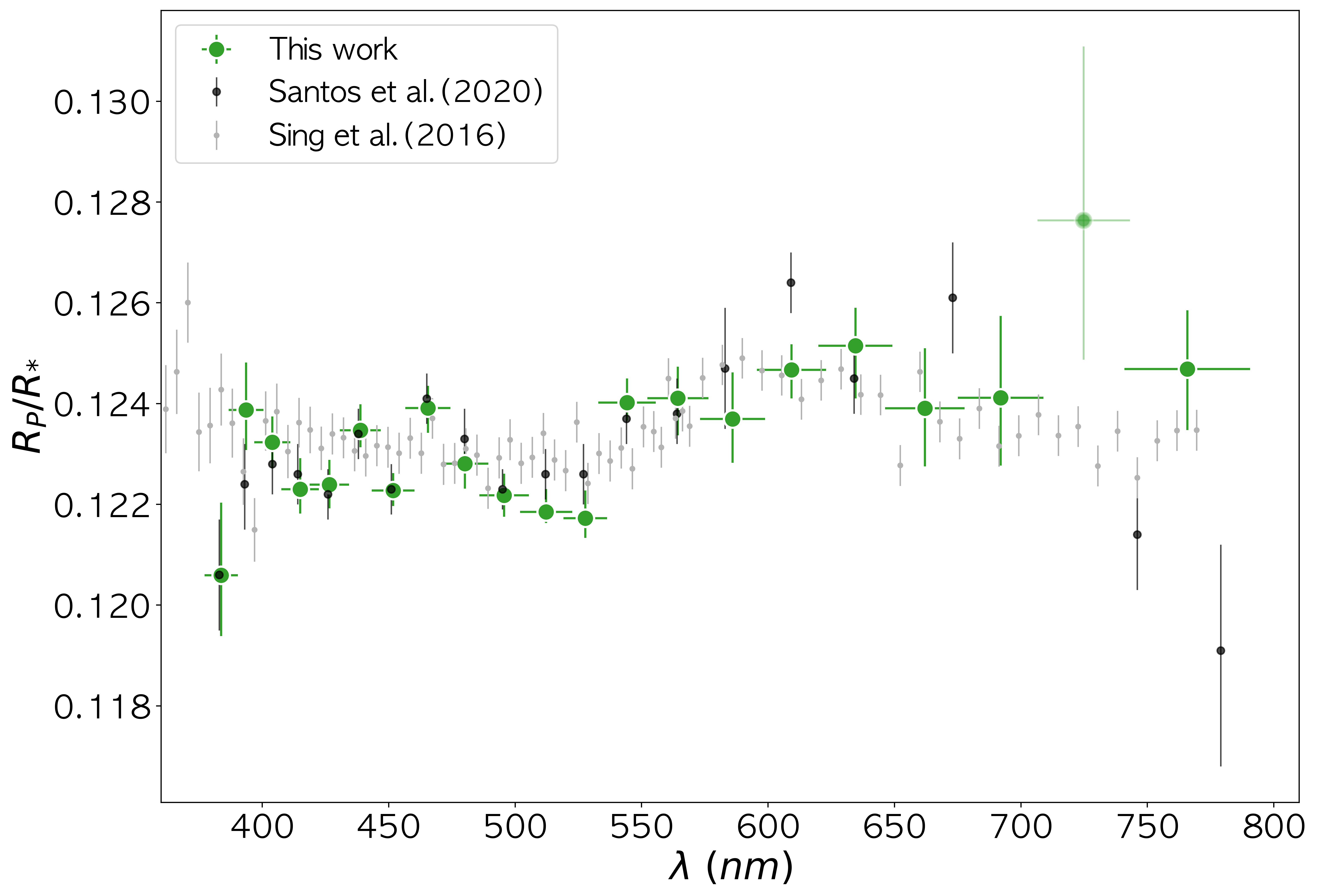}
    \caption{Comparison between the transmission spectra of HD~209458b obtained by \citet{Santos-20} (black circle), by \citet{Sing-16} (gray circle),and from our analysis (green circle). The x-axis error bars stand for the width of the wavelength bin, while the y-axis error bars stand for the uncertainties of the estimated planet-star radius ratio in each wavelength bin taken as the $1\sigma$ values of the posterior distributions from the fitting. We applied an offset to the values of the radius ratio obtained from our analysis, subtracting the average difference between our values and the \citet{Santos-20} values. Similarly, we introduced an offset to the \cite{Sing-16} data set to subtract the average difference with the \citet{Santos-20} values.}
    \label{fig:Transm_spectra}
\end{figure}

The transmission spectrum of HD~209458b, resulting from our fit to 21 wavelength bins and combining the SKYSUB data from two nights, is shown in Fig.~\ref{fig:Transm_spectra} (see values in Table~\ref{tab:wavelength_correspondance_and_results}). The non-SKYSUB subtracted data results are given in  Fig.~\ref{fig:Skysub_comparison}. The retrieved transmission spectra from individual transits agree well with one another and with the combined nights (Fig.~\ref{fig:Separate_nights}). Our data are the same as used in \citet{Santos-20}, although they employed the chromatic RM method on the combined two nights to retrieve the broadband transmission spectra of HD~209458b. Figure~\ref{fig:Transm_spectra} shows a comparison of our results with those presented in \cite{Santos-20}, and with the transmission spectrum obtained from STIS multiband photometry data from the Hubble Space Telescope (HST) \citep{Sing-16}.

Our retrieved transmission spectrum agrees well with both the \citet{Santos-20} and \cite{Sing-16} results. Moreover, the achieved precision on the planet radius in each wavelength bin is compatible with the ones obtained from HST and from ESPRESSO combining two transits. In particular, the comparison between the transmission spectrum obtained through our analysis and the spectrum presented by \cite{Santos-20} reveals identical patterns. We find a mostly flat spectrum with a strong increase in radius above  550~nm, which is consistent with both comparison spectra. Our study, however, shows less scatter than  \cite{Santos-20}, especially above 550~nm, where the retrieved spectrum is more in line with \cite{Sing-16}.

Two spectral bins, centered at 384~nm and 725~nm, deviate noticeably from the general trend of the spectrum. The 384~nm bin contains the spectral slices from the bluest edge of the ESPRESSO and has a low S/N. The retrieved transmission spectrum by \cite{Santos-20} exhibits a similar deviation in this first bin. Regarding the 725~nm bin, we identify a notable deviation compared to the \cite{Sing-16} spectrum, which is probably caused by the strong telluric contamination by $H_2 O$ and $O_2$ in the last two bins of our spectra \citep{Tellurics_synth_atlas}. \cite{Santos-20} were not able to obtain transmission data in this bin.

\subsection{Atmospheric retrieval}
\label{sec:retrieval}

We performed spectral retrieval analyses on the ESPRESSO transmission spectrum (21 wavelength bins). The model consists of two components:
\begin{equation}
D_{\lambda,\mathrm{obs}}=D_{\lambda,\mathrm{true}}\times\epsilon_{\lambda},
\end{equation}
where $D_{\lambda,\mathrm{true}}$ comes from the planetary atmosphere and $\epsilon_{\lambda}$ refers to the stellar contamination.

The forward model of the planetary atmosphere was implemented by PLATON \citep[PLanetary Atmospheric Transmission for Observer Noobs,][]{2019PASP..131c4501Z,2020ApJ...899...27Z}, which is parameterized by: the planet radius at 1~bar ($R_\mathrm{p,1bar}$), an isothermal temperature ($T_\mathrm{iso}$), the cloud-top pressure ($P_\mathrm{cloud}$) for the assumption of uniform clouds, an enhancement factor over the nominal H$_2$ Rayleigh scattering ($A_\mathrm{RS}$), and the volume mixing ratio for a given species other than H, H$_2$, and He ($X_i$). The atmosphere was assumed to be dominated by hydrogen and helium, with a solar composition of H, H$_2$, and He. Fixed values were adopted in the forward model for the planet mass ($M_\mathrm{p}=0.714$~M$_\mathrm{J}$) and the stellar radius ($R_\mathrm{*}=1.162$~R$_\sun$) from TEPCat \citep[Transiting Extrasolar Planets CATalogue,][]{2011MNRAS.417.2166S}.

The stellar contamination \citep{2014ApJ...791...55M,2018ApJ...853..122R,2019AJ....157...96R} included the contribution from both spots and faculae, which is parameterized by the spot temperature ($T_\mathrm{spot}$), the spot coverage ($f_\mathrm{spot}$), the facula temperature ($T_\mathrm{facu}$), and the facula coverage ($f_\mathrm{facu}$), in the form of 
\begin{equation}
\epsilon_{\lambda}=\frac{1}{1-f_\mathrm{spot}\Bigg(1-\frac{S_{\lambda,T_\mathrm{spot}}}{S_{\lambda,T_\mathrm{phot}}}\Bigg)-f_\mathrm{facu}\Bigg(1-\frac{S_{\lambda,T_\mathrm{facu}}}{S_{\lambda,T_\mathrm{phot}}}\Bigg)},
\end{equation}
where the spectrum $S_{\lambda,T}$ was interpolated in the BT-NextGen (AGSS2009) stellar spectral grid \citep{2012RSPTA.370.2765A} provided in PLATON. The photosphere was assumed to have a temperature of $T_\mathrm{phot}=6118$~K.

\begin{figure*}
\centering
\includegraphics[width=\textwidth]{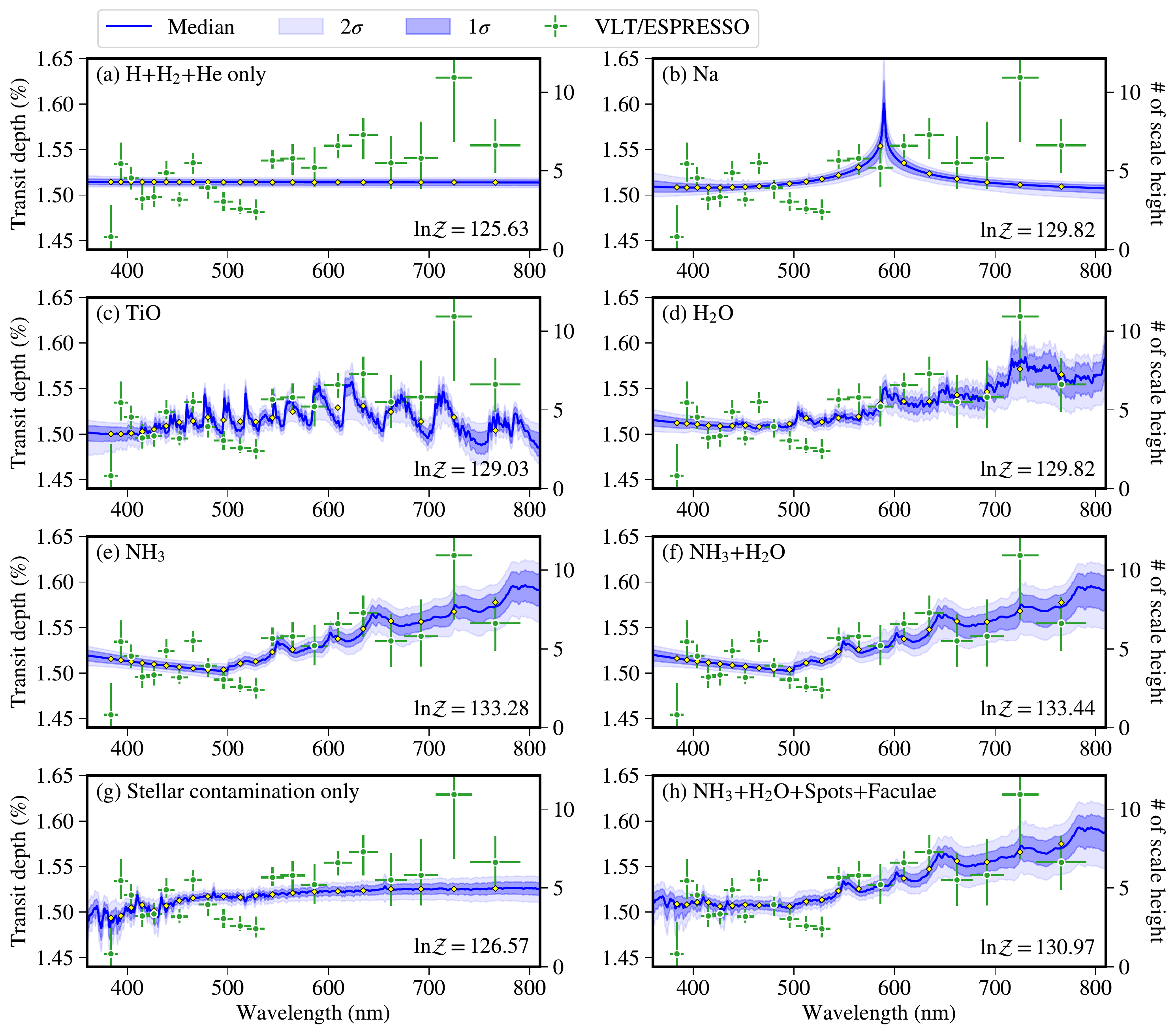}
\caption{Comparison between the ESPRESSO transmission spectrum and the retrieval analyses under different model assumptions. Panel (a) considers a planetary atmosphere composed of only H, H$_2$, and He. Additional gas absorbers are added in Panels (b)-(f). Panel (g) assumes only the stellar contamination to account for any transit depth variation, while Panel (h) includes both planetary atmosphere and stellar contamination. Of these model assumptions, an atmosphere containing H$_2$O and NH$_3$ gives the highest Bayesian evidence (Panel f).}
\label{fig:retrieval}
\end{figure*}

To explain the ESPRESSO transmission spectrum, we considered three scenarios: (I) a purely planetary atmosphere (i.e., $\epsilon_{\lambda}=1$), (II) a purely stellar contamination (i.e., $D_{\lambda,\mathrm{true}}=D_0$), and (III) a combination of both planetary atmosphere and stellar contamination. We used \texttt{PyMultiNest} \citep{2014A&A...564A.125B} to implement the multimodal nested sampling to explore the posteriors, which also gives the Bayesian evidence ($\mathcal{Z}$). A total of 1,000 live points were adopted in each retrieval analysis, resulting in a typical uncertainty of $\sim$0.07--0.10 for $\ln\mathcal{Z}$. Table \ref{tab:retrieval_evidence} summarizes the resulting $\ln\mathcal{Z}$ of various model assumptions. Figure \ref{fig:retrieval} shows a selection of those listed in Table \ref{tab:retrieval_evidence}. Table \ref{tab:retrieval_param} presents the representative prior setup and posterior estimates for scenario III. The criteria of \citet{kass_raftery_1995} were adopted to interpret the Bayes factor ($B_{10}=\mathcal{Z}_1/\mathcal{Z}_0$) of two hypotheses, where $\Delta\ln\mathcal{Z}=1, 3, 5$ are considered as the starting points of positive, strong, and very strong, respectively. 

\begin{figure*}
\centering
\includegraphics[width=\textwidth]{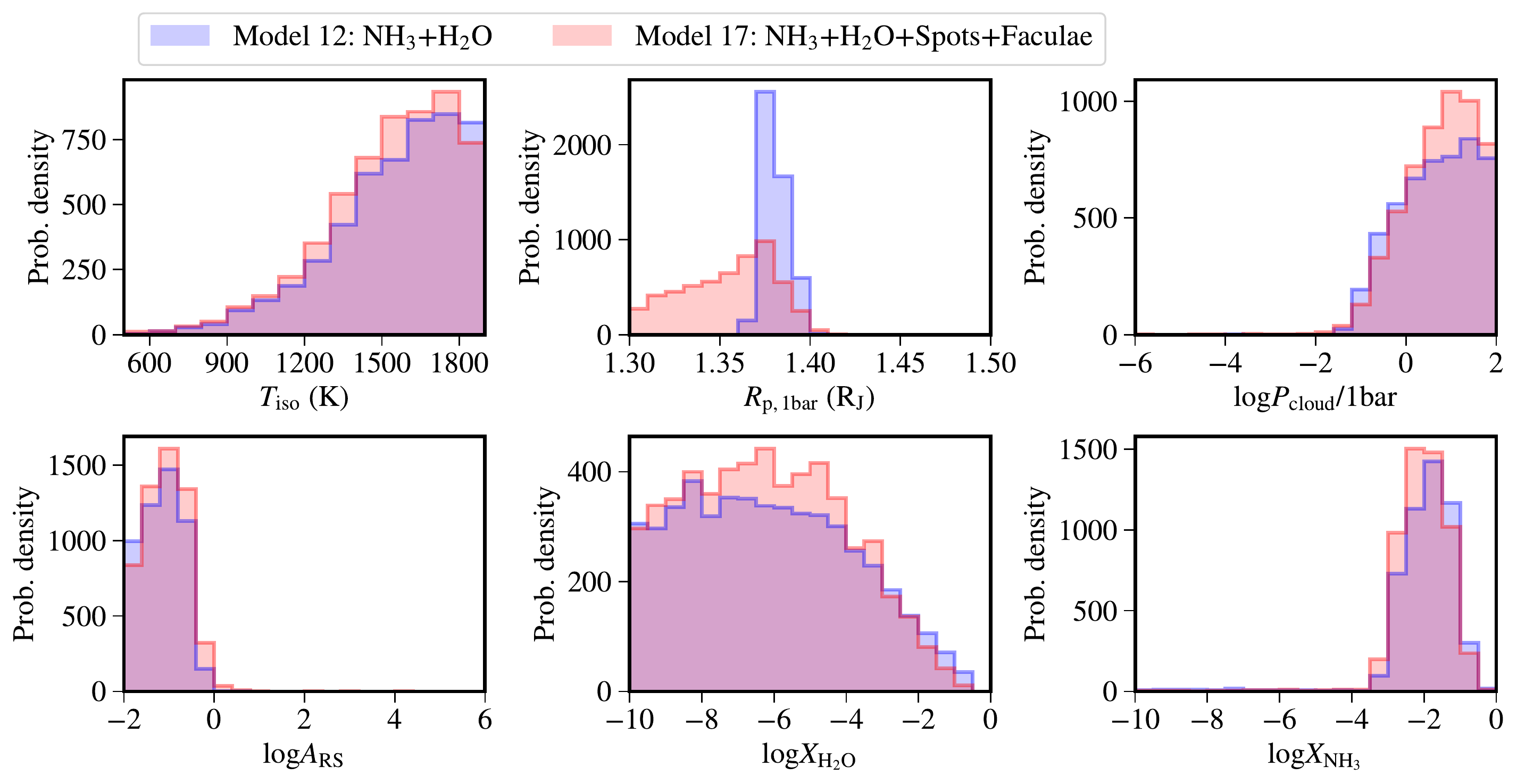}
\caption{Posterior distributions of the common atmospheric parameters in Model 12 and Model 17.}
\label{fig:retrieval_posteriors}
\end{figure*}

In the scenario of a purely planetary atmosphere, the model assumption of no additional absorbers other than H, H$_2$, and He is considered as a reference since it could be a flat line or a Rayleigh scattering slope, depending on the altitude of the cloud deck. To the base of H, H$_2$, and He, a variety of additional optical absorbers, including Na, K, TiO, VO, H$_2$O, and NH$_3$, were then added to the atmosphere to explore the contribution of the observed spectral signature. The models with additional optical absorbers are always strongly favored over the reference model, except for VO at a positive level. The most favored model assumption is the model with NH$_3$ or with both NH$_3$ and H$_2$O. In contrast, the ESPRESSO transmission spectrum cannot be explained by the scenario of purely stellar contamination. Finally, we investigated the scenario of a combination of both planetary atmosphere and stellar contamination, where the component of planetary atmosphere included additional absorbers of both NH$_3$ and H$_2$O. This model assumption (Model 17) decreases the Bayesian evidence by $\Delta\ln\mathcal{Z}=2.5$ compared to the scenario of a purely planetary atmosphere (Model 12), indicating that the data are positively against the inclusion of stellar contamination in the model assumption. Nevertheless, the inclusion of stellar contamination barely changes the retrieved atmospheric parameters (see Fig.~\ref{fig:retrieval_posteriors}).

Furthermore, we performed the same retrieval analyses on the transmission spectrum without subtracting an offset of 0.005 and on the transmission spectrum without the 725~nm bin. In both cases, the model selection results remain the same, that is, the models with NH$_3$ are always favored by our data. The resulting posteriors also remain consistent for all the parameters except for $R_\mathrm{p,1bar}$, which would be larger if the offset were not subtracted. In addition to the assumption of uniform clouds, we also experimented with the assumption of patchy clouds with an additional free parameter to account for the cloud coverage. However, the assumption of patchy clouds is strongly disfavored by our data according to the decreasing model evidence, as shown in Table \ref{tab:retrieval_evidence}.

\citet{Santos-20} found that Na, TiO, or both were favored by their version of the ESPRESSO transmission spectrum. By investigating more optical absorbers, our retrieval analyses indicate that NH$_3$ is likely present in the atmosphere of HD 209458b. On the other hand, in the presence of NH$_3$, it is not possible to distinguish whether or not Na, K, or H$_2$O exists, while TiO and VO are strongly disfavored. 

Our potential inference of NH$_3$ is consistent with recent studies on the same planet. \citet{2017MNRAS.469.1979M} reported the first possible inference of NH$_3$ using the low-resolution transmission spectrum acquired with the Wide Field Camera 3 (WFC3) aboard the HST. Newly, using the high-resolution cross-correlation technique, \citet{Giacobbe2021} robustly detected NH$_3$ in the transmission spectrum; H$_2$O, CO, HCN, CH$_4$, and C$_2$H$_2$ were also detected with the near-infrared echelle spectrograph GIANO-B mounted at the 3.6~m Telescopio Nazionale Galileo.

Recently, \cite{CasasayasBarris_2021} explored the atmosphere of \hdton, also using these same ESPRESSO observations. They did not detect Na absorption in the high-resolution transmission spectrum of this planet around single lines, due to the overlap of the Doppler shadow and the position of the expected atmospheric absorption. Our best model is not conclusive regarding the presence of Na.

\begin{table*}
\centering
\caption{Bayesian evidence for various model assumptions obtained from three versions of the transmission spectrum.}
\label{tab:retrieval_evidence}
\begin{tabular}{cl|cc|cc|cc}
\hline\hline\noalign{\smallskip}
\# & Model assumption & \multicolumn{2}{c}{Full w/ offset} & \multicolumn{2}{c}{Full w/o offset} & \multicolumn{2}{c}{w/o 725nm \& w/ offset}\\ \noalign{\smallskip}
  &                   & $\ln\mathcal{Z}$ & $\Delta\ln\mathcal{Z}$ & $\ln\mathcal{Z}$ & $\Delta\ln\mathcal{Z}$ & $\ln\mathcal{Z}$ & $\Delta\ln\mathcal{Z}$\\ \noalign{\smallskip}
\hline\noalign{\smallskip}
\multicolumn{8}{c}{\it Planetary atmosphere (uniform clouds)} \\ \noalign{\smallskip}
1 & No additional absorbers   & 125.63 & ref. & 124.90 & ref. & 120.75 & ref.\\ \noalign{\smallskip}
2 & Na                        & 129.82 & 4.2  & 129.02 & 4.1 & 125.07 & 4.3\\ \noalign{\smallskip}
3 & K                         & 128.95 & 3.3  & 127.85 & 2.9 & 123.37 & 2.6\\ \noalign{\smallskip}
4 & TiO                       & 129.03 & 3.4  & 128.18 & 3.3 & 123.96 & 3.2\\ \noalign{\smallskip}
5 & VO                        & 127.14 & 1.5  & 126.57 & 1.7 & 122.09 & 1.3\\ \noalign{\smallskip}
6 & H$_2$O                    & 129.82 & 4.2  & 129.33 & 4.4 & 123.77 & 3.0\\ \noalign{\smallskip}
7 & NH$_3$                    & 133.28 & 7.7  & 132.58 & 7.7 & 127.44 & 6.7\\ \noalign{\smallskip}
8 & NH$_3$+Na                 & 132.77 & 7.1  & 132.02 & 7.1 & 127.15 & 6.4\\ \noalign{\smallskip}
9 & NH$_3$+K                  & 132.87 & 7.2  & 132.13 & 7.2 & 126.95 & 6.2\\ \noalign{\smallskip}
10 & NH$_3$+TiO               & 129.79 & 4.2  & 128.83 & 3.9 & 124.60 & 3.8\\ \noalign{\smallskip}
11 & NH$_3$+VO                & 129.27 & 3.6  & 128.76 & 3.9 & 123.84 & 3.1\\ \noalign{\smallskip}
12 & NH$_3$+H$_2$O            & 133.44 & 7.8  & 132.61 & 7.7 & 127.36 & 6.6\\ \noalign{\smallskip}
13 & NH$_3$+H$_2$O+Na+K       & 132.65 & 7.0  & 131.91 & 7.0 & 126.94 & 6.2\\ \noalign{\smallskip}
14 & NH$_3$+H$_2$O+Na+K+TiO+VO & 129.11 & 3.5 & 128.91 & 3.4 & 123.79 & 3.0\\ \noalign{\smallskip}
15 & Na+TiO                   & 129.23 & 3.6  & 128.38 & 3.5 & 124.27 & 3.5\\ \noalign{\smallskip}
\hline \noalign{\smallskip}
\multicolumn{8}{c}{\it Stellar contamination} \\ \noalign{\smallskip}
16 & Spots+Faculae            & 126.57 & 0.9  & 125.92 & 1.0 & 121.28 & 0.5\\ \noalign{\smallskip}
\hline \noalign{\smallskip}
\multicolumn{8}{c}{\it Planetary atmosphere (uniform clouds) and stellar contamination combined} \\ \noalign{\smallskip}
17 & NH$_3$+H$_2$O+Spots+Faculae & 130.97 & 5.3 & 130.50 & 5.6 & 124.69 & 3.9\\ \noalign{\smallskip}
\hline \noalign{\smallskip}
\multicolumn{8}{c}{\it Planetary atmosphere (patchy clouds)} \\ \noalign{\smallskip}
18 & NH$_3$        & 129.29 & 3.7 & 128.85 & 4.0 & 123.80 & 3.1 \\ \noalign{\smallskip}
19 & NH$_3$+H$_2$O & 129.59 & 4.0 & 129.37 & 4.5 & 123.73 & 3.0 \\ \noalign{\smallskip}
\hline \noalign{\smallskip}
\multicolumn{8}{c}{\it Planetary atmosphere (patchy clouds) and stellar contamination combined} \\ \noalign{\smallskip}
20 & NH$_3$+H$_2$O+Spots+Faculae & 128.91 & 3.3 & 128.94 & 4.0 & 123.23 & 2.5\\ \noalign{\smallskip}
\hline\noalign{\smallskip}
\end{tabular}
\end{table*}

\begin{table}
\centering
\caption{Parameter priors and posteriors for the spectral retrievals.}
\label{tab:retrieval_param}
\begin{tabular}{lcr}
\hline\hline\noalign{\smallskip}
Parameter & Prior & Posterior \\ \noalign{\smallskip}
\hline\noalign{\smallskip}
\multicolumn{3}{c}{Planetary atmosphere} \\ \noalign{\smallskip}
$R_\mathrm{p,1bar}$ (R$_\mathrm{J}$) & $\mathcal{U}(1.3,1.5)$ & $1.36 ^{+0.021}_{-0.034}$\\ \noalign{\smallskip}
$T_\mathrm{iso}$ (K) & $\mathcal{U}(500,1900)$ & $1576 ^{+207}_{-293}$\\ \noalign{\smallskip}
$\log P_\mathrm{cloud}/\mathrm{1bar}$ & $\mathcal{U}(-6,2)$ & $0.8 ^{+0.7}_{-0.9}$\\ \noalign{\smallskip}
$\log A_\mathrm{RS}$ & $\mathcal{U}(-2,6)$ & $-1.1 ^{+0.5}_{-0.5}$\\ \noalign{\smallskip}
$\log X_{\mathrm{H}_2\mathrm{O}}$ & $\mathcal{U}(-10,0)$ & $-6.3 ^{+2.5}_{-2.4}$\\ \noalign{\smallskip}
$\log X_{\mathrm{N}\mathrm{H}_3}$ & $\mathcal{U}(-10,0)$ & $-2.0 ^{+0.7}_{-0.7}$\\ \noalign{\smallskip}
\hline \noalign{\smallskip}
\multicolumn{3}{c}{Stellar contamination} \\ \noalign{\smallskip}
$T_\mathrm{spot}$ (K) & $\mathcal{U}(2000,5918)$ & $3312 ^{+1365}_{-806}$\\ \noalign{\smallskip}
$f_\mathrm{spot}$ (\%) & $\mathcal{U}(0,100)$ & $5 ^{+5}_{-3}$\\ \noalign{\smallskip}
$T_\mathrm{facu}$ (K) & $\mathcal{U}(6118,7000)$ & $6193 ^{+122}_{-55}$\\ \noalign{\smallskip}
$f_\mathrm{facu}$ (\%) & $\mathcal{U}(0,100)$ & $14 ^{+25}_{-10}$\\ \noalign{\smallskip}
\hline\noalign{\smallskip}
\end{tabular}
\tablefoot{In the spectral retrieval analyses, the model assumption could be purely planetary atmosphere, or purely stellar contamination, or a combination of both. The table presents the posterior estimates for the model assumption where both planetary atmosphere (with NH$_3$ and H$_2$O) and stellar contamination are considered, i.e., Model 17 in Table \ref{tab:retrieval_evidence}.}
\end{table}

\section{Discussion and conclusions}\label{sec:Conclusion}

In this work we have developed a novel chromatic line-profile tomography technique, named CHOCOLATE (CHrOmatiC line prOfiLe tomogrAphy TEchnique), to retrieve the broadband transmission spectrum of transiting exoplanets using high-resolution spectroscopy. The CHOCOLATE method utilizes the \texttt{SOAP} code to produce physically-motivated Doppler tomography models in different wavelength bins; it then uses these \texttt{SOAP} models to fit the observational data via an MCMC fitting procedure. We validated its performance on an ESPRESSO data set of the well-known and well-studied exoplanet HD~209458b.

The transmission spectrum of HD~209458b obtained through CHOCOLATE agrees extremely well with previous results \citep{Santos-20,Sing-16} that were derived using different methods and data, thus demonstrating its usefulness and reliability. We have shown that our methodology achieves similar precision (even better in some bins) than the different methodologies (i.e., chromatic RM) using the same data \citep{Santos-20}. We have explored different scenarios using PLATON and found that a planetary atmosphere model containing H$_2$O and NH$_3$ is the preferred scenario, as derived from the spectral retrieval on the transmission spectrum, with a confidence of $\Delta\ln\mathcal{Z}=7.8$.

In our study, we use a version of SOAP that assumes CCFs to be pure Gaussian, while in reality stellar CCFs have a so-called C shape due to the stellar granulation \citep{Dravins_1981,SOAP_Dumusque}. Additionally, removing or masking strong spectral lines known to be affected by short-term variability in the ESPRESSO 1D spectra could further improve the stability of the CCFs, especially on the redder orders. Probing the impact of these differences is beyond scope of the current paper but will be pursued in forthcoming works.

The application of the CHOCOLATE methodology is especially interesting for future studies on exoplanets around moderate and fast rotating stars, where the chromatic RM method cannot be applied due to the fact that the RV precision is dominated by the rotational broadening of the star. Moreover, for planets around young and very active host stars, where the CCF residual matrix is heavily affected by the presence of stellar active regions \citep{Palle-20}, CHOCOLATE could assist by more efficiently modeling and retrieving their transmission spectra, taking \texttt{SOAP}'s ability to model stellar active regions into account.

There has been a recent discussion in the literature about the great potential of simultaneously exploring exoplanet atmospheres at low and high spectral resolution \citep{Brogi2017,2018A&A...612A..53P}. This is because high-resolution transmission spectroscopy probes the cores of spectral lines generated in the planetary thermosphere and exosphere, including those of escaping material, while low-resolution spectroscopy is more sensitive to the higher planetary troposphere and can capture broader spectral features, such as molecular bands, clouds and hazes, or Rayleigh scattering processes. Our results demonstrate how it will be possible, using only high dispersion spectroscopy, to retrieve the broad transmission spectrum of a planet, which had previously been thought to only be possible via low-resolution observations. This will be particularly important in the era of the extremely large telescopes (ELTs). Given their small field of view and a general lack of suitable comparison stars, low-resolution spectroscopy will generally not be possible. High-resolution spectroscopy with fiber-fed ultra-stable spectrographs, via the use of CHOCOLATE and similar techniques, will be able to probe both the lines profiles and the broad continuum features of exoplanet transmission spectra, providing a complete picture of their atmospheres.

\begin{acknowledgements}
     
     We thank the anonymous referee for insightful suggestions, which added the clarity of this paper.
     E. E-B. acknowledges financial support from the European Union and the State Agency of Investigation of the Spanish Ministry of Science and Innovation (MICINN) under the grant PRE2020-093107 of the Pre-Doc Program for the Training of Doctors (FPI-SO) through FEDER, FSE and FDCAN funds.
     E. E-B. also wants to acknowledge funding from the Instituto de Astrofísica de Canarias (IAC) Summer Grant in Astrophysics 2020.
     This work is partly financed by the Spanish Ministry of Economics and Competitiveness through grants PGC2018-098153-B-C31. The ESPRESSO Instrument Project was partially funded through SNSF’s FLARE Programme for large infrastructures. This work made use of PyAstronomy.
     G.C. acknowledges the support by the National Natural Science Foundation of China (Grant No. 42075122, 12122308), the Natural Science Foundation of Jiangsu Province (Grant No.\,BK20190110), and Youth Innovation Promotion Association CAS (2021315).
     This work was supported by FCT - Fundação para a Ciência e a Tecnologia through national funds and by FEDER through COMPETE2020 - Programa Operacional Competitividade e Internacionalização by these grants: UID/FIS/04434/2019; UIDB/04434/2020; UIDP/04434/2020; PTDC/FIS-AST/32113/2017 \& POCI-01-0145-FEDER-032113; PTDC/FIS-AST/28953/2017 \& POCI-01-0145-FEDER-028953; PTDC/FIS-AST/28987/2017 \& POCI-01-0145-FEDER-028987.
     This work has been carried out within the framework of the National Centre of Competence in Research PlanetS supported by the Swiss National Science Foundation. This project has received funding from the European Research Council (ERC) under the European Union’s Horizon 2020 research and innovation programme (project Four Aces grant agreement No 724427).
     We acknowledge funding from the European Research Council under the European Union’s Horizon 2020 research and innovation program under grant agreement No 694513.
     G. M. has received funding from the European Union’s Horizon 2020 research and innovation programme under the Marie Skłodowska-Curie grant agreement No. 895525.
     R. A. is a Trottier Postdoctoral Fellow and acknowledges support from the Trottier Family Foundation. This work was supported in part through a grant from FRQNT. This work has been carried out within the framework of the National Centre of Competence in Research PlanetS supported by the Swiss National Science Foundation. The authors acknowledge the financial support of the SNSF.
     This work has been carried out in the frame of the National Centre for Competence in Research ``PlanetS'' supported by the Swiss National Science Foundation (SNSF). This project has received funding from the European Research Council (ERC) under the European Union's Horizon 2020 research and innovation programme (project {\sc Spice Dune}, grant agreement No 947634).
     J.L-B. acknowledges financial support received from ”la Caixa” Foundation (ID 100010434) and from the European Union’s Horizon 2020 research and innovation programme under the Marie Skłodowska-Curie grant agreement No 847648, with fellowship code LCF/BQ/PI20/11760023.
     O.D.S.D. is supported in the form of work contract (DL 57/2016/CP1364/CT0004) funded by national funds through Fundação para a Ciência e Tecnologia (FCT).
     V.A. acknowledges the support from FCT through Investigador FCT contract nr. IF/00650/2015/CP1273/CT0001.
     A.S.M. acknowledges financial support from the Spanish Ministry of Science and Innovation (MICINN) under the 2019 Juan de la Cierva Programme.
     A.S.M., J.I.G.H., C.A.P. and R.R. acknowledge financial support from the Spanish Ministry of Science and Innovation (MICINN) through projects AYA2017-86389-P.
     J.I.G.H. also acknowledges financial support from the Spanish MICINN under 2013 Ram\'on y Cajal program RYC-2013-14875.
     M.R.Z.O. acknowledges financial support through project PID2019-109522G8-C51.
     The INAF authors acknowledge financial support of the Italian Ministry of Education, University, and Research with PRIN 201278X4FL and the "Progetti Premiali" funding scheme.

\end{acknowledgements}


\bibliographystyle{aa}
\bibliography{HD209}

\begin{appendix}
\section{Wavelength bin definition and numeric results}

We specify in Table~\ref{tab:wavelength_correspondance_and_results} the exact correspondence between the defined bins and the range of wavelengths covered by each bin, which comes as an outcome of \emph{Step~2} in the data reduction (Sect.~\ref{sec:data_reduction}). We also show in Table~\ref{tab:wavelength_correspondance_and_results} the estimated values for the radius and the limb-darkening coefficients from the fitting procedure.

\begin{table}[htb]
    \centering
    \caption{Central wavelength, wavelength range covered by each defined bin and planet-star radius ratio ($R_{P}/R_{*}$) values estimated through the fitting procedure for each wavelength bin using the SKYSUB data set (shown in Fig.~\ref{fig:Transm_spectra}). We note that a $-~0.005$ offset has been introduced to the resulting $R_{P}/R_{*}$ values to subtract the average difference with the \cite{Santos-20} values. Each bin contains eight ESPRESSO slices, except the last one, which contains ten. Moreover, it should be noted that this wavelength binning definition does not involve spectral slice overlap, but a slight overlap in wavelength is present between adjacent bins as a consequence of the way orders are created by dispersion plus cross-dispersion, as seen on ESPRESSO and similar echelle spectrographs.}
    \begin{tabular}{l c c c}
        \hline \hline \\
         & \textit{Central} & \textit{Wavelength} & \\
         Bin & \textit{Wavelength} & \textit{Range} & \textit{$R_{P}/R_{*}$} \\
          & \textit{(nm)} & \textit{(nm)} &  \\[1pt]
         \hline \\
         1 & 383.75 & 377.15 - 390.34 & $0.1206_{- 0.0012}^{+ 0.0014}$ \\[4pt]
         2 & 393.62 & 386.76 - 400.48 & $0.1239_{- 0.0008}^{+ 0.0009}$ \\[4pt]
         3 & 404.01 & 396.87 - 411.15 & $0.1232_{- 0.0005}^{+ 0.0005}$ \\[4pt]
         4 & 414.98 & 407.53 - 422.42 & $0.1223_{- 0.0005}^{+ 0.0006}$ \\[4pt]
         5 & 426.55 & 418.78 - 434.31 & $0.1224_{- 0.0005}^{+ 0.0005}$ \\[4pt]
         6 & 438.78 & 430.67 - 446.89 & $0.1235_{- 0.0003}^{+ 0.0005}$ \\[4pt]
         7 & 451.74 & 443.26 - 460.22 & $0.1223_{- 0.0003}^{+ 0.0003}$ \\[4pt]
         8 & 465.49 & 456.60 - 474.38 & $0.1239_{- 0.0005}^{+ 0.0004}$ \\[4pt]
         9 & 480.10 & 470.77 - 489.42 & $0.1228_{- 0.0005}^{+ 0.0003}$ \\[4pt]
         10 & 495.65 & 485.85 - 505.45 & $0.1222_{- 0.0004}^{+ 0.0004}$ \\[4pt]
         11 & 512.26 & 501.94 - 522.57 & $0.1219_{- 0.0002}^{+ 0.0005}$ \\[4pt]
         12 & 527.73 & 519.13 - 536.33 & $0.1217_{- 0.0004}^{+ 0.0005}$ \\[4pt]
         13 & 544.27 & 532.89 - 555.65 & $0.1240_{- 0.0004}^{+ 0.0005}$ \\[4pt]
         14 & 564.35 & 552.27 - 576.42 & $0.1241_{- 0.0007}^{+ 0.0006}$ \\[4pt]
         15 & 585.96 & 573.12 - 598.80 & $0.1237_{- 0.0009}^{+ 0.0009}$ \\[4pt]
         16 & 609.30 & 595.60 - 622.99 & $0.1247_{- 0.0006}^{+ 0.0005}$ \\[4pt]
         17 & 634.57 & 619.92 - 649.22 & $0.1251_{- 0.0010}^{+ 0.0008}$ \\[4pt]
         18 & 662.03 & 646.31 - 677.74 & $0.1239_{- 0.0012}^{+ 0.0012}$ \\[4pt]
         19 & 691.97 & 675.04 - 708.89 & $0.1241_{- 0.0013}^{+ 0.0016}$ \\[4pt]
         20 & 724.75 & 706.46 - 743.04 & $0.1276_{- 0.0028}^{+ 0.0035}$ \\[4pt]
         21 & 765.80 & 740.95 - 790.64 & $0.1247_{- 0.0012}^{+ 0.0012}$ \\[4pt]
         \hline
    \end{tabular}
    \label{tab:wavelength_correspondance_and_results}
\end{table}

\section{Noise estimation}

\begin{figure}[htb]
    \centering
    \includegraphics[width=1.\linewidth]{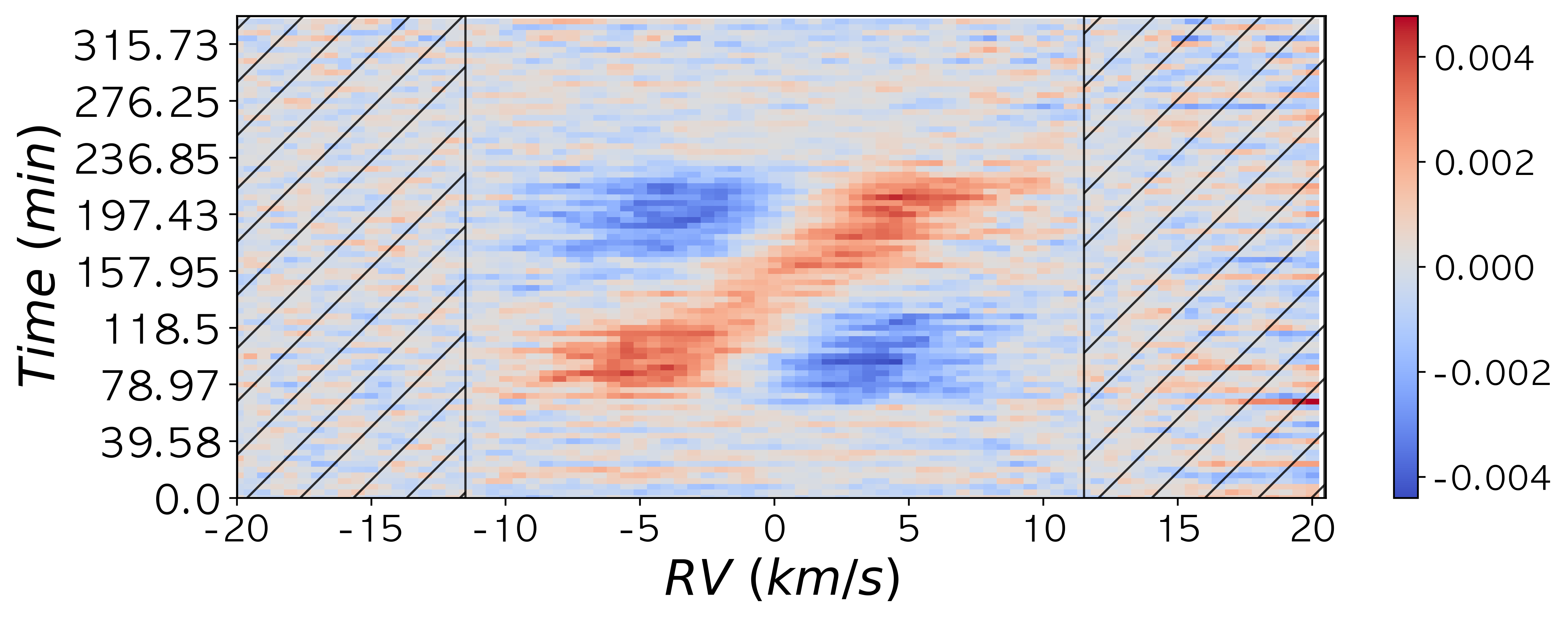}
    \caption{Regions used for the noise estimation (line-filled rectangles) marked over the processed data matrix using the 426~nm-centered bin as an example.}
    \label{noise_example}
\end{figure}

We show in Fig.~\ref{noise_example} the CCF residual matrix regions used to calculate the standard deviation for the noise estimation.

\section{\texttt{SOAP} CCF fitting}

\begin{figure}[htb]
    \centering
    \includegraphics[width=1.\linewidth]{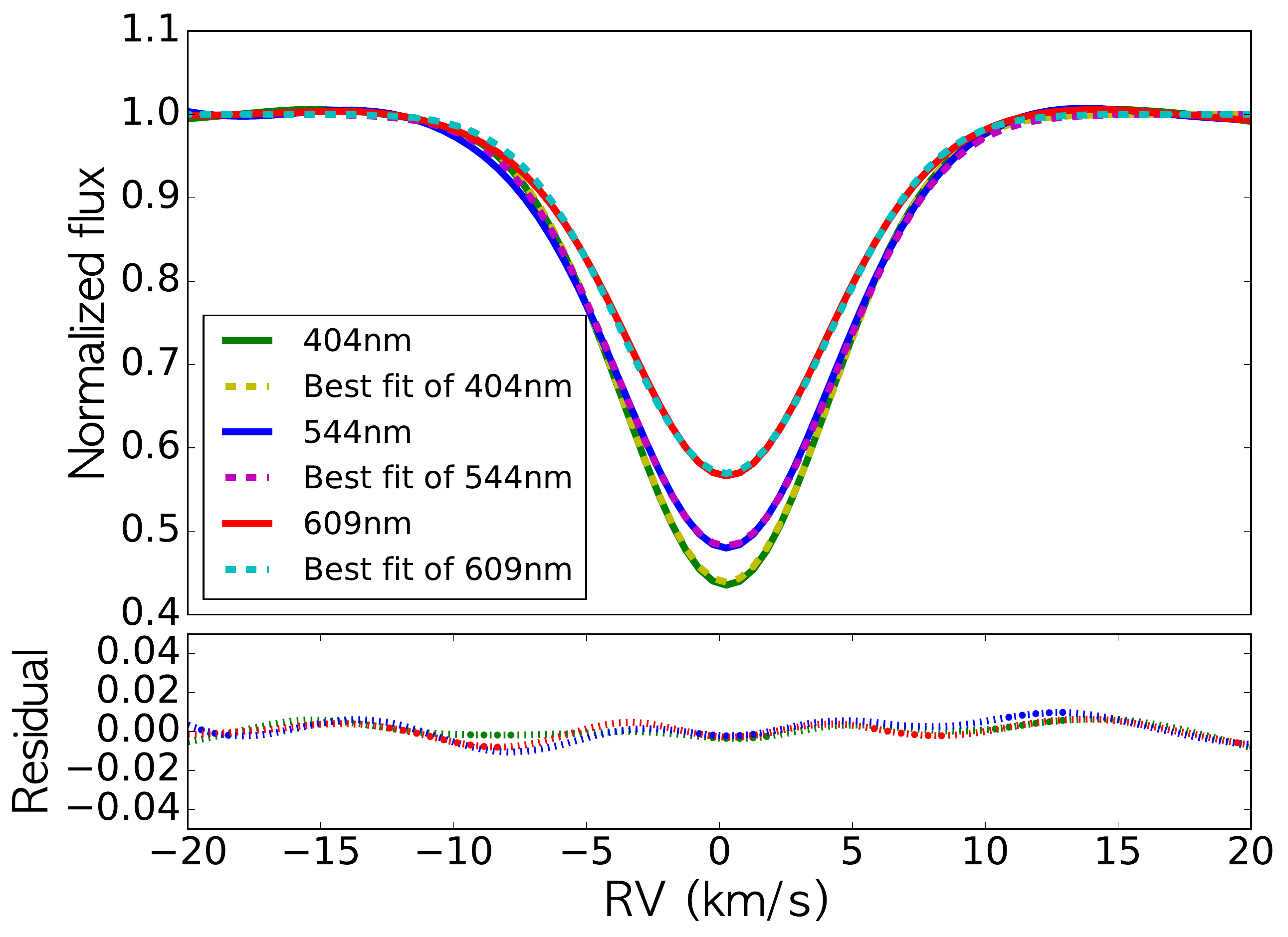}
    \caption{Example of the best fitted \texttt{SOAP} CCF to the master CCF in different wavelength bins, which leads to the determination of the Gaussian FWHM and depth parameters. The solid lines represent the master CCFs from the observations in the 440~nm (green), 544~nm (blue), and 609~nm (red) bins, respectively. The \texttt{SOAP} CCF resulting from the best-fit Gaussian parameters is shown by the dashed lines.}
    \label{fig: sigmaprof_fit}
\end{figure}

In this appendix we show the procedure used to fit \texttt{SOAP}'s CCF to the master CCF in each wavelength bin and obtain the FWHM and depth values of the Gaussian from the best fit (Fig.~\ref{fig: sigmaprof_fit}). The resulting values are fixed in \texttt{SOAP} configuration for each wavelength bin before the emcee-based fitting procedure used to determine $R_{P}/R_{*}$, $u_{1}$, and $u_{2}$ is executed.

\section{Posterior distributions}

\begin{figure}[htb]
    \centering
    \includegraphics[width=1.\linewidth]{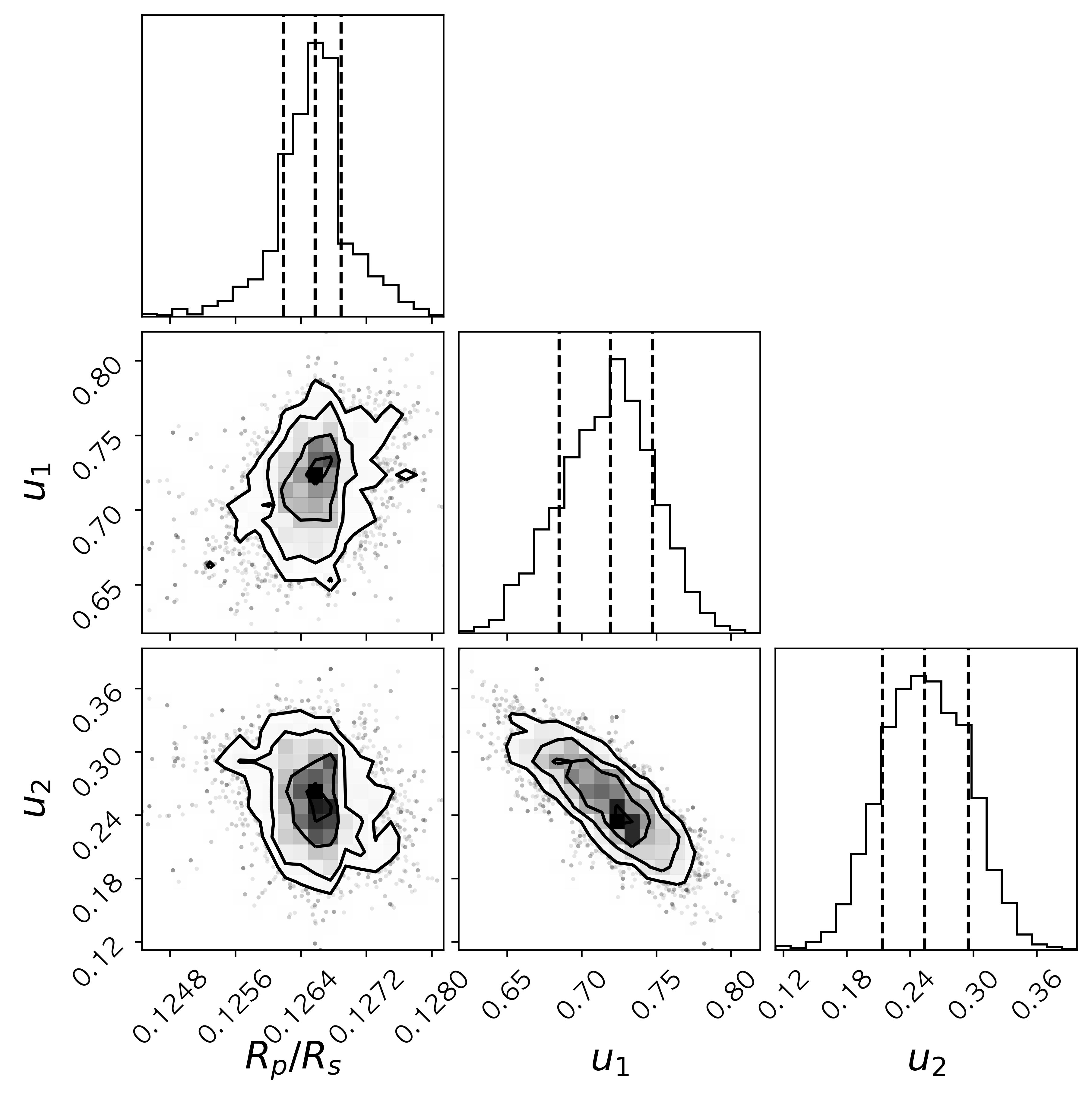}
    \caption{Posterior distributions obtained through the fitting procedure using the $528~nm$-centered bin as an example. The central dashed line indicates the best-fit value of each fitting parameter. The surrounding dashed lines mark the $1\sigma$ values of each distribution, which were taken to be the value enclosed in the 68.3\% of the posterior distributions.}
    \label{CornerPlotExample}
\end{figure}

An example of the posterior distributions of the free parameters ($R_{P}/R_{*}$, $u_{1}$, $u_{2}$) resulting from the \texttt{emcee}-based fitting procedure is shown in Fig.~\ref{CornerPlotExample}.

\section{Comparison with non-SKYSUB data}\label{App:SKYSUB_comparison}
\begin{figure}[h!]
    \centering
    \includegraphics[width=1.\linewidth]{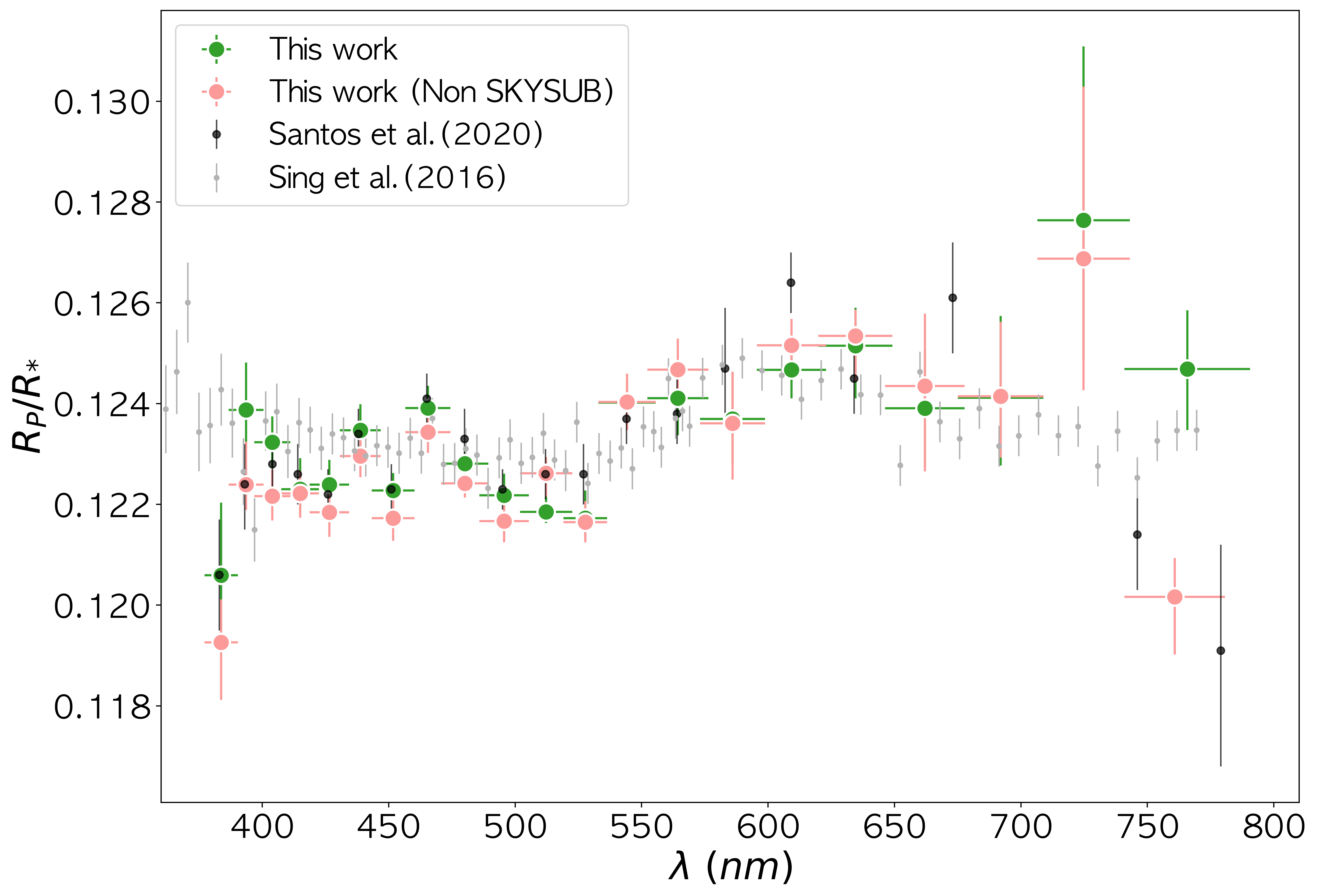}
    \caption{Comparison between transmission spectra of \hdton obtained from SKYSUB and non-SKYSUB data sets.}
    \label{fig:Skysub_comparison}
\end{figure}

In the results section we present the transmission spectra obtained from the ESPRESSO SKYSUB data set extracted by the DRS pipeline. However, we also tested the CHOCOLATE methodology on the non-SKYSUB ESPRESSO data set (Fig.~\ref{fig:Skysub_comparison}) to evaluate the impact of sky subtraction on the results. Consequently, we found that there is no meaningful impact on the results. Nevertheless, we obtained that the non-SKYSUB data set performed spuriously on the last bin.

\section{Results from separate nights}

\begin{figure}[h!]
    \centering
    \includegraphics[width=1.\linewidth]{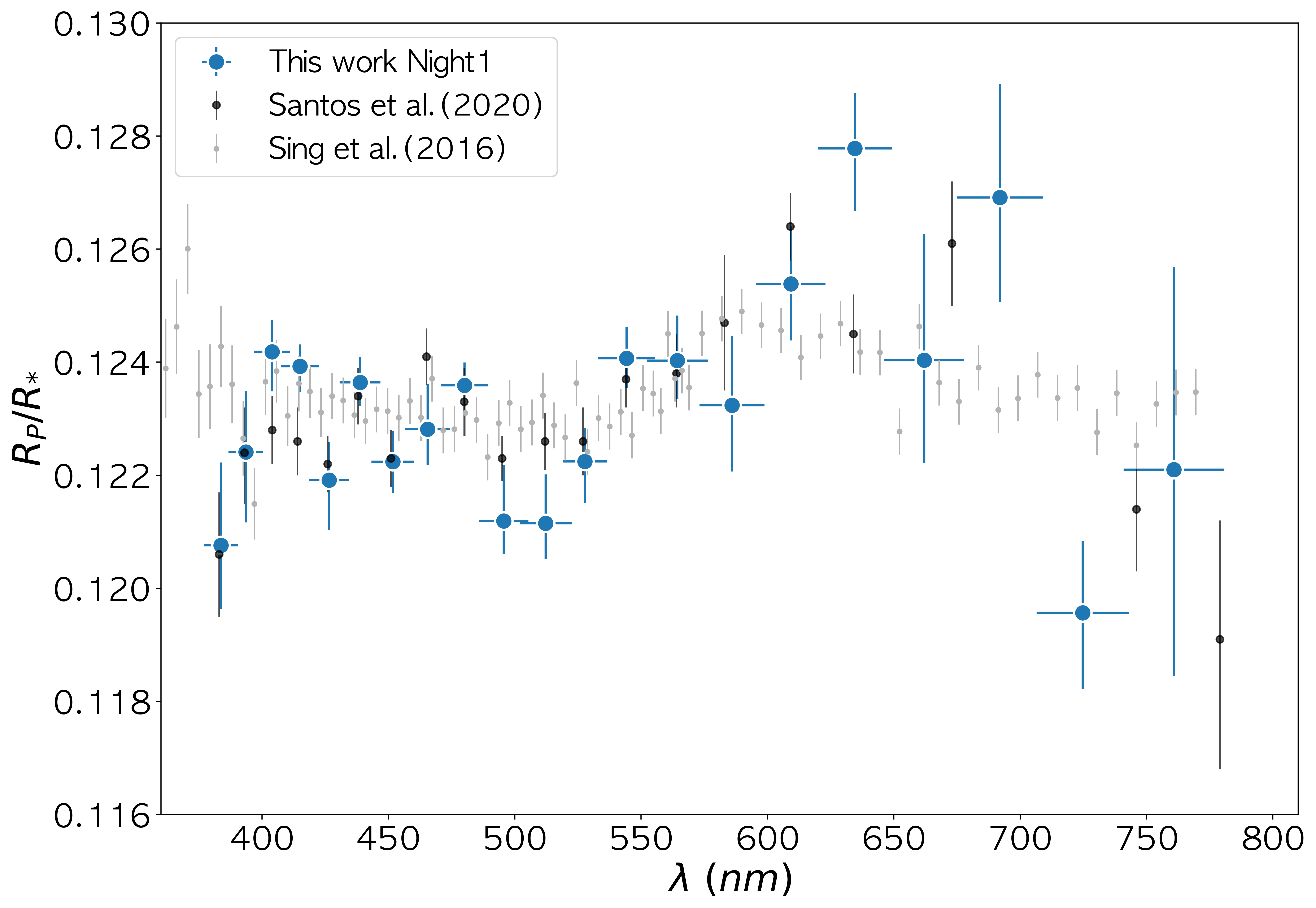}
    \includegraphics[width=1.\linewidth]{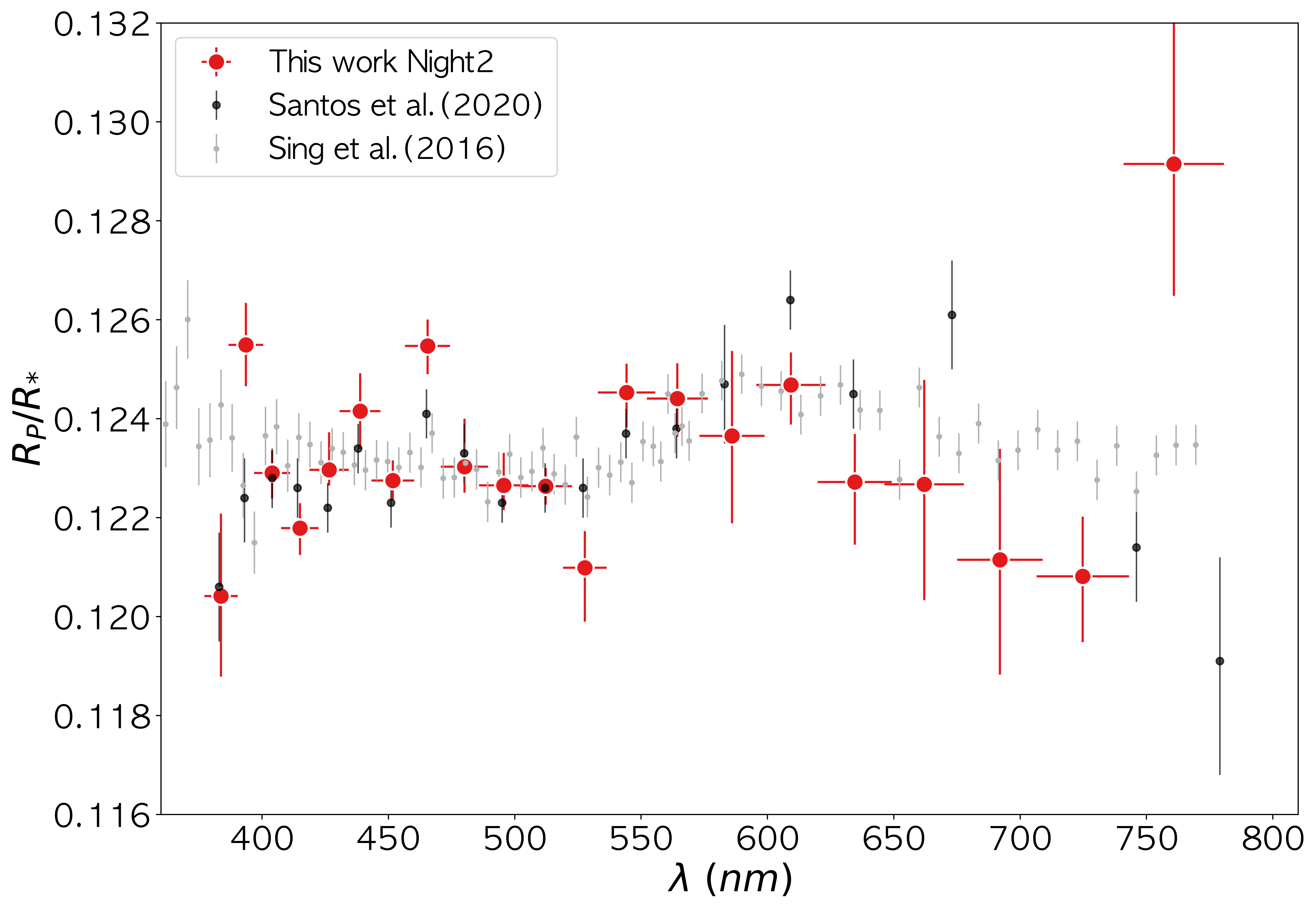}
    \includegraphics[width=1.\linewidth]{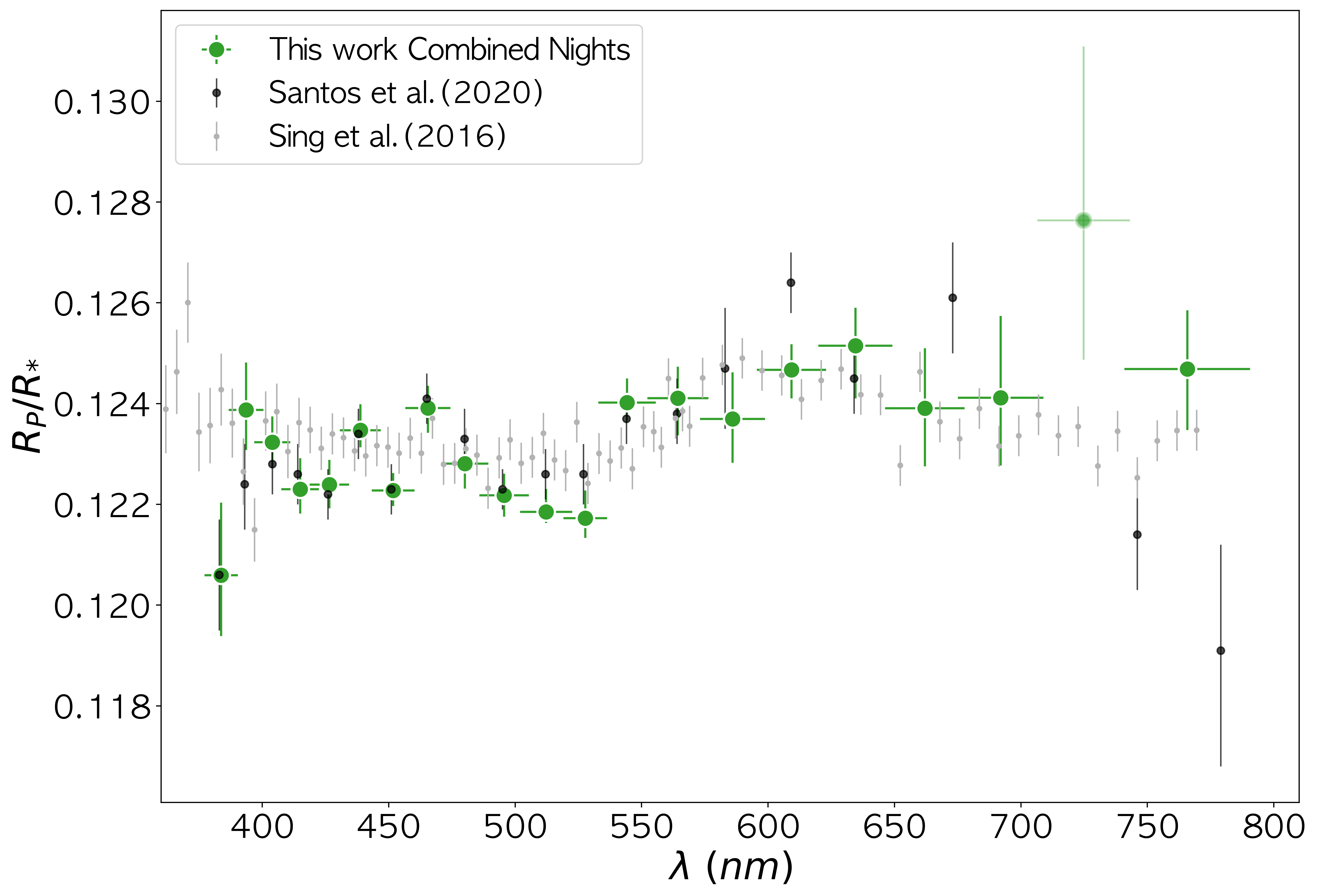}
    \caption{Transmission spectra of \hdton obtained from our analysis using Night 1 (top panel, blue circle), Night 2 (middle panel, red circle) and combined (bottom panel, green circle) data sets. Each result is compared to the transmission spectra obtained by \cite{Santos-20} (black circle) and \cite{Sing-16} (gray circle).}
    \label{fig:Separate_nights}
\end{figure}

We show in Fig.~\ref{fig:Separate_nights} the transmission spectra obtained by applying our analysis independently to Night~1 and Night~2 data sets, and we compare them with the results of our analysis obtained from the combined data set presented in Sect.~\ref{sec:Results}. We find that in the case of HD209458~b, which is a quiet star, the combination of two observing nights is good enough to obtain a conclusive result. However, in the case of an active star, where the activity may differ between nights, a larger combination of data sets might be needed.

\end{appendix}

\end{document}